\documentclass[twocolumn]{aastex631}
\usepackage{amsmath}
\usepackage{mathrsfs}
\usepackage{graphicx}
\usepackage{xcolor}
\usepackage{dcolumn}
\usepackage{bm}
\usepackage{amssymb}
\usepackage{amsmath}
\usepackage{verbatim}
\usepackage[utf8]{inputenc}
\usepackage{tikz,hyperref}
\usepackage{dcolumn}

\definecolor{darkyellow}{rgb}{0.545098,0,0}

\shortauthors{Zhang, Yu, Yang, \& Nadler}

\begin{document}

\title{The GD-1 stellar stream perturber as a core-collapsed self-interacting dark matter halo}

\author[0009-0009-2791-1684]{Xingyu Zhang}
\email{zhang-xy19@mails.tsinghua.edu.cn}
\affiliation{Department of Physics, Tsinghua University, Beijing 100084, China}
\affiliation{Department of Physics and Astronomy, University of California, Riverside, California 92521, USA}

\author[0000-0002-8421-8597]{Hai-Bo Yu}
\email{haiboyu@ucr.edu}
\affiliation{Department of Physics and Astronomy, University of California, Riverside, California 92521, USA}

\author[0000-0002-5421-3138]{Daneng Yang }
\email{danengy@ucr.edu}
\affiliation{Department of Physics and Astronomy, University of California, Riverside, California 92521, USA}
\affiliation{Purple Mountain Observatory, Chinese Academy of Sciences, Nanjing 210023, China}

\author[0000-0002-1182-3825]{Ethan O.~Nadler}
\email{enadler@carnegiescience.edu}
\affiliation{Carnegie Observatories, 813 Santa Barbara Street, Pasadena, CA 91101, USA}
\affiliation{Department of Physics \& Astronomy, University of Southern California, Los Angeles, CA 90007, USA}
\affiliation{Department of Astronomy \& Astrophysics, University of California, San Diego, La Jolla, CA 92093, USA}

\begin{abstract}
The GD-1 stellar stream exhibits spur and gap structures that may result from a close encounter with a dense substructure. When interpreted as a dark matter subhalo, the perturber is denser than predicted in the standard cold dark matter (CDM) model. In self-interacting dark matter (SIDM), however, a halo could evolve into a phase of gravothermal collapse, resulting in a higher central density than its CDM counterpart. We conduct high-resolution controlled N-body simulations to show that a collapsed SIDM halo could account for the GD-1 perturber's high density. We model a progenitor halo with a mass of $3\times10^8~M_\odot$, motivated by a cosmological simulation of a Milky Way analog, and evolve it in the Milky Way's tidal field. For a cross section per mass of $\sigma/m\approx30\textup{--}100~{\rm cm^2~g^{-1}}$ at $V_{\rm max }\sim10~{\rm km~s^{-1}}$, the enclosed mass of the SIDM halo within the inner $10~{\rm pc}$ can be increased by more than an order of magnitude compared to its CDM counterpart, leading to a good agreement with the properties of the GD-1 perturber. Our findings indicate that stellar streams provide a novel probe into the self-interacting nature of dark matter.

\end{abstract}

\keywords{
\href{http://astrothesaurus.org/uat/353}{Dark matter (353)};
\href{http://astrothesaurus.org/uat/1880}{Galaxy dark matter halos (1880)};
\href{http://astrothesaurus.org/uat/2166}{Stellar streams (622)}
}

\section{Introduction}
\label{sec:intro}

Stellar streams form when globular clusters or dwarf galaxies are tidally stripped. There are more than $100$ streams discovered in the Milky Way, see, e.g.,~\cite{Bonaca:2024dgc,S5:2021ooa,DES:2018imd} and references therein. Among them, the GD-1 stream is one of the longest and coldest streams~\citep{Grillmair:2006bd}, and it has been used to constrain the Milky Way's gravitational potential~\citep{Koposov:2009hn-cons, Bovy:2016chl-cons, Bonaca:2018-cons, Malhan:2019-cons}. The GD-1 stream has rich structural properties, such as the gaps~\citep[e.g.][]{Carlberg:2013gxa,deBoer:2018-gap,deBoer:2020,Banik:2019cza,Malhan:2022} and spur~\citep{Price:2018-gap,Bonaca:2018fek-main,Bonaca:2020psc}, suggesting that it has been perturbed through interactions with a substructure in the Milky Way.

In particular,~\cite{Bonaca:2018fek-main} demonstrated that the perturber must be surprisingly dense to account for the spur and gap features in the GD-1 stream. Assuming a Hernquist density profile, the perturber's mass is estimated to be in the range of  $10^{5.5}\textup{--}10^8~M_\odot$, with a scale radius of $\lesssim20~{\rm pc}$; recent encounters within the last $1~{\rm Gyr}$ are favored. The perturber is significantly denser than the subhalos predicted in the standard cold dark matter (CDM) model, at the $\sim3\sigma$ level. Thus even if the perturber were a known satellite galaxy of the Milky Way, its unusually high density would remain puzzling. Furthermore, none of the known globular clusters can match the orbit of the inferred perturber~\citep{Bonaca:2018fek-main,Doke:2022jro-gaia3}. 

In this work, we assume that the GD-1 perturber is a dark matter subhalo and explore its formation in within the framework of self-interacting dark matter (SIDM), see~\cite{Tulin:2017ara-core,Adhikari:2022sbh} for reviews and references therein. The gravothermal evolution of an SIDM halo occurs in two sequential phases. In the core-forming phase, dark matter self-interactions transport heat inward, resulting in a shallow density core, while in the core-collapsing phase, heat transfer reverses, leading to a higher central density than in the CDM counterpart~\citep[e.g.][]{Balberg:2002ue-CC,Koda:2011yb-CC,Essig:2018pzq-tc,Feng:2020kxv}. Notably, SIDM models with large cross sections could explain the high density of the strong lensing perturber for SDSSJ0946+1006~\citep{Nadler:2023nrd-tid,Minor:2020hic,Vegetti:2009cz} {\it and} the low density of the Crater~II satellite galaxy~\citep{Zhang:2024ggu,Borukhovetskaya:2021ahz}, both challenging CDM. It is intriguing to explore the SIDM scenario to account for the high density of the GD-1 perturber.

We will analyze progenitors of CDM subhalos from a zoom-in cosmological simulation of a Milky Way analog from~\cite{Yang:2022mxl-tid-cosmo-CDM,DES:2019ltu-Nadler2020}, and explicitly show that their inner densities are systematically lower than those inferred for the GD-1 perturber. We then take one of the progenitor halos, with a mass of $\sim10^8~M_\odot$, and evolve it in the tidal field of the Milky Way, including both halo and stellar components. For a self-interacting cross section in the range $\sigma/m=30\textup{--}100~{\rm cm^2~g^{-1}}$ at $V_{\rm max}\sim10~{\rm km~s^{-1}}$, the SIDM halo enters the collapse phase within $3\textup{--}6~{\rm Gyr}$ while evolving in the tidal field. By the final snapshot, its enclosed mass within the inner $10~{\rm pc}$ is increased by more than an order of magnitude compared to its CDM counterpart, making it consistent with the high density of the GD-1 perturber. Additionally, we will discuss future investigations aimed at further improvement.

The rest of this Letter is organized as follows: In Section~\ref{sec:cdm}, we discuss the properties of CDM halos in the cosmological zoom-in simulation of a Milky Way analog. In Section~\ref{sec:setup}, we introduce the setup of our N-body simulations. In Section~\ref{sec:results}, we present the properties of our simulated SIDM and CDM subhalos and compare them with the GD-1 perturber. In Section~\ref{sec:diss}, we discuss future investigations for further improvement and conclude. In Appendix~\ref{sec:iso}, we present the SIDM simulation of an isolated halo for testing numerical artifacts that could lead to violation of energy conservation. In Appendix~\ref{sec:con}, we show the convergence test.

\section{CDM halos of a Milky Way analog}
\label{sec:cdm}

\begin{figure*}[t]
	\centering
	\includegraphics[scale=0.22]{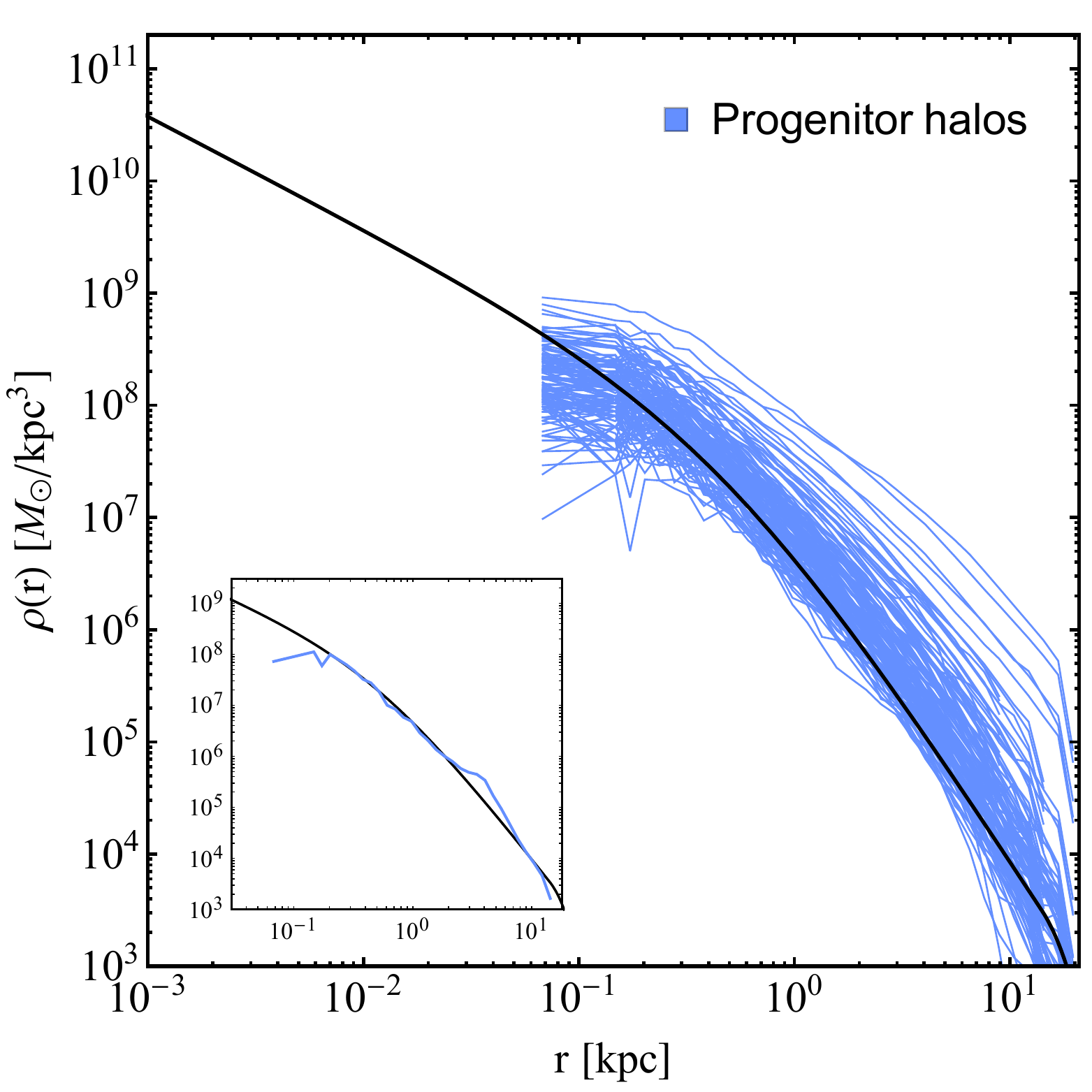}~~
	\includegraphics[scale=0.22]{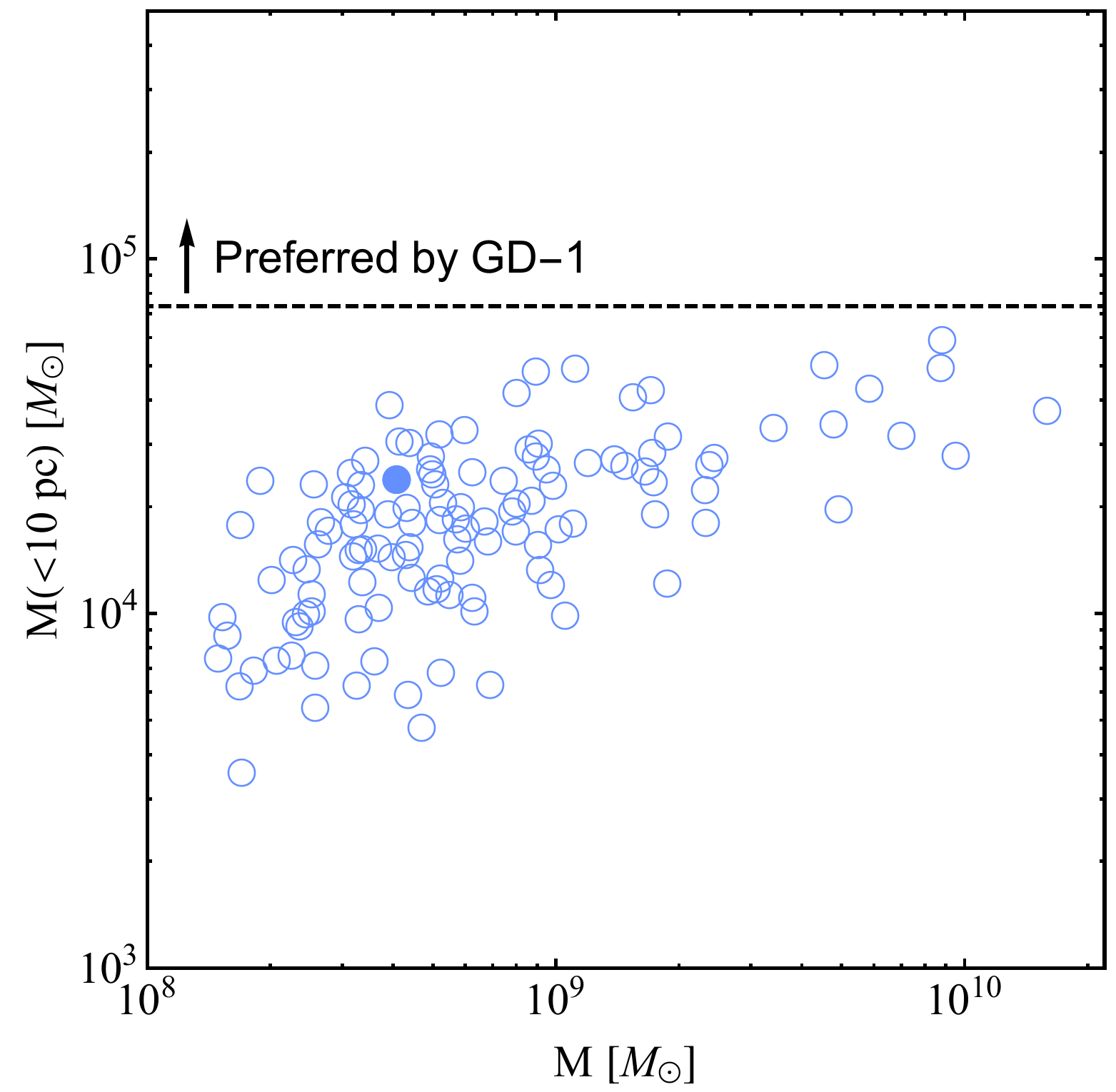}~~
	\includegraphics[scale=0.22]{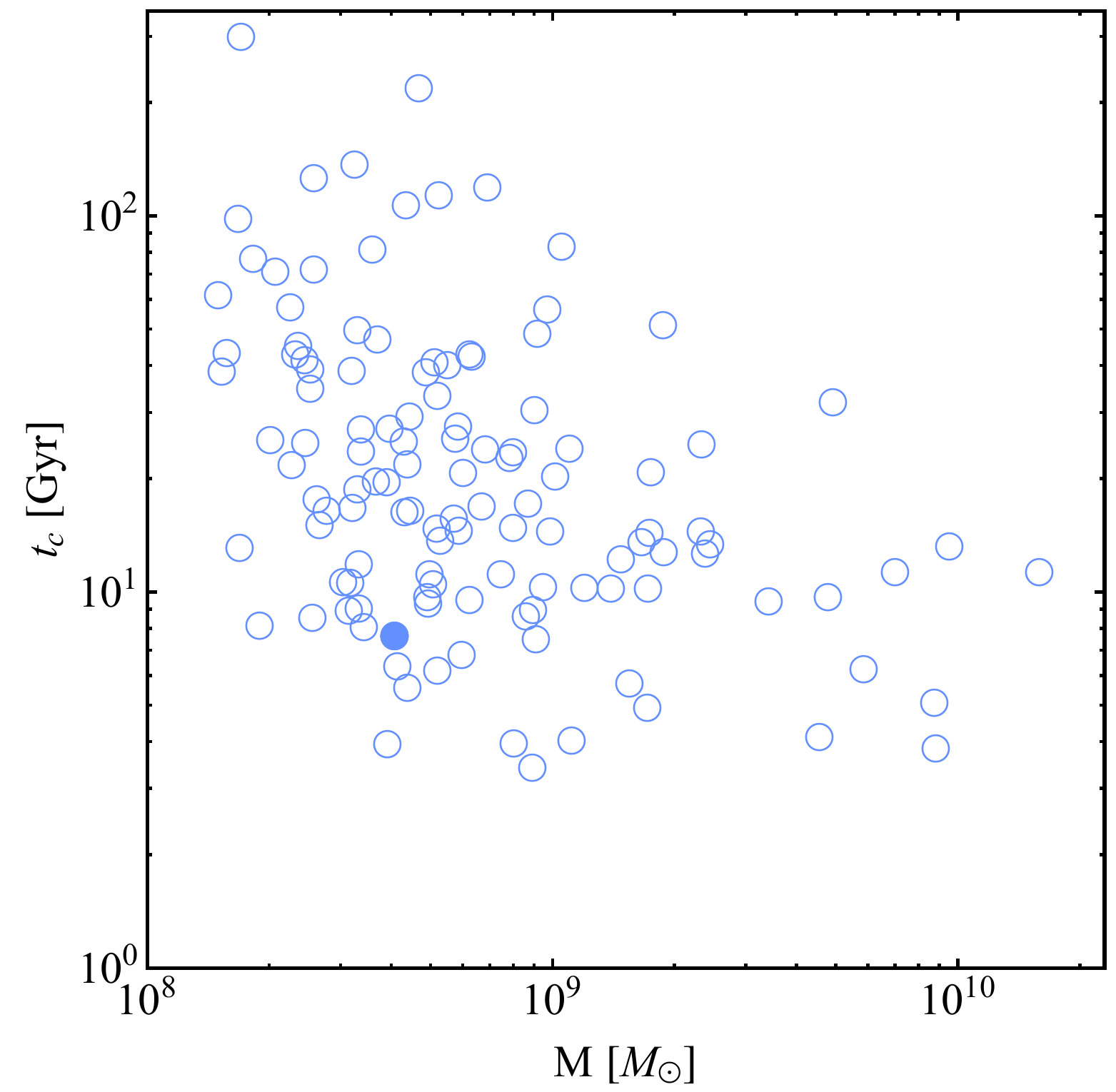}
	\caption{{\bf Left:} density profiles for $125$ CDM progenitor halos at their infall times from the zoom-in cosmological simulation of a Milky Way analog in~\cite{Yang:2022mxl-tid-cosmo-CDM} (blue). The black line indicates an NFW density profile fitted to one of the progenitor halos (see the inset panel), which will be used as the initial condition for our controlled SIDM and CDM simulations. 
	{\bf Middle:} enclosed mass within inner $10~{\rm pc}$ vs virial mass for the CDM progenitor halos. The filled circle marks the halo used for the initial condition. The horizontal dashed gray line indicates the inner mass within $10~{\rm pc}$ for a reference Hernquist profile with the scale radius $r_{\rm H} = 15~{\rm pc}$ and the total mass $M = 4.6 \times 10^5~M_\odot$, representing one of the least dense perturber models for the GD-1 stellar stream in~\cite{Bonaca:2018fek-main}. 
		{\bf Right:} estimated SIDM collapse timescale vs virial mass for the progenitor halos, assuming $\sigma/m=50~{\rm cm^2~g^{-1}}$. The filled circle marks the halo used for the initial condition in our controlled simulations.}
	\label{fig:cdm}
\end{figure*}

We first present progenitor halos from a cosmological zoom-in CDM-only simulation of a Milky Way analog~\citep{Yang:2022mxl-tid-cosmo-CDM,DES:2019ltu-Nadler2020}, with initial conditions drawn from the suite in~\cite{Mao:2015yua}. This simulated system includes a main halo with a mass of $1.14\times10^{12}~M_\odot~h^{-1}\approx1.6\times10^{12}~M_\odot~(h=0.7)$, and a Large Magellanic Cloud analog. The simulation has a particle mass of $4\times 10^4~M_\odot~h^{-1}$, a Plummer-equivalent softening length of $\epsilon = 0.08 ~ {\rm kpc}~h^{-1}$, and a spline length of $\ell = 2.8 \epsilon = 0.22 ~ {\rm kpc}~h^{-1}$, the characteristic length scale of the smoothing kernel used to calculate gravitational forces between particles~\citep{Springel:2000yr-GADGET2,Springel:2005mi-GADGET2}.

We select subhalos of the main halo with the virial mass larger than $10^{8}~M_\odot~h^{-1}$ at $z=0$, and then identify their progenitors at infall. With the mass cut, there will be at least $2500$ simulation particles for each progenitor halo so that we can accurately reconstruct its density profile. The radial resolution of the cosmological simulation $\ell\approx0.3~{\rm kpc}$ is more than an order of magnitude larger than the radial scale relevant for the GD-1 perturber. To overcome this resolution limit, we fit each progenitor halo with a truncated Navarro-Frenk-White (NFW) profile~\citep{Errani:2020wgn} for the region $r>0.3~{\rm kpc}$,
\begin{equation}
\rho(r)=\frac{\rho_s}{\left(r / r_s\right)\left(1+r / r_s\right)^2} \times \frac{ \exp( -r/r_{\rm cut} ) }{(1+r_s/r_{\rm cut})^{0.3}},
\label{eq:nfw}
\end{equation}
where $\rho_s$ and $r_s$ are the scale density and radius, respectively, and $r_{\rm cut}$ is the truncation radius due to tidal stripping. We determine the three parameters for each progenitor at infall, achieving excellent overall fit quality. The truncated NFW profile is then extrapolated inward to compute the total enclosed mass within $r=10~{\rm pc}$. Additionally, we confirm that many progenitor halos can be well-fitted with the standard NFW profile, while some exhibit density profiles slightly steeper than $r^{-3}$ in the outer regions. For these cases, the standard NFW fit may introduce bias and overestimate the central density. However, the truncated NFW profile provides a significantly better fit.

Figure~\ref{fig:cdm} (left) shows the density profiles for the $125$ progenitor halos at infall (blue). We also present the fit to one of the progenitors (black), which will be used as the initial condition for our SIDM simulations; see the detailed comparison in the inset panel. For this halo, the standard NFW profile provides a good fit. The simulated density profile is flattened for $r\lesssim0.3~{\rm kpc}$ due to the resolution limit. However, we expect that the NFW profile provides a good approximation for extrapolating the density inward before the halo undergoes significant tidal stripping. 

Figure~\ref{fig:cdm} (middle) shows the enclosed mass within $10~{\rm pc}$ vs virial mass of the progenitor halos at infall. For comparison, we include a reference case from the viable parameter region of the GD-1 perturber in~\citealt{Bonaca:2018fek-main} (their Figure 6): a Hernquist scale radius of $r_{\rm H} = 15~{\rm pc}$ and a total mass of $M= 4.6\times 10^5~M_\odot$, which approximately corresponds to a substructure with the minimum density required to explain the spur and gap features of the GD-1 stream. For this reference case, the enclosed mass within $10~{\rm pc}$ is $\approx7.4\times10^4~M_\odot$, as denoted by the horizontal line in the middle panel. We see that none of the CDM progenitor halos are sufficiently dense to be the perturber, and this conclusion holds when comparing the enclosed mass within $r=15~{\rm pc}$. The inner density of these CDM halos would further decrease as they evolve within the Milky Way's tidal field. The progenitor CDM halos shown in Figure~\ref{fig:cdm} correspond to subhalos with masses $>10^8~M_\odot~h^{-1}$ at $z=0$. We plan to relax this mass threshold and examine halos with lower masses. Based on the resolution limit in the cosmological CDM simulation~\citep{Yang:2022mxl-tid-cosmo-CDM}, we expect to reconstruct the density profiles of progenitors for subhalos with masses a few times $10^7~M_\odot~h^{-1}$ using the truncated NFW profile in Equation~\ref{eq:nfw}. A more detailed investigation will be deferred to future work.

For the CDM progenitor halos, we estimate the timescale of gravothermal collapse in SIDM~\citep{Pollack:2014rja-tc, Essig:2018pzq-tc}
\begin{equation}
	t_c=\frac{150}{C} \frac{1}{r_s \rho_s\left(\sigma_{\mathrm{eff}} / m\right)} \frac{1}{\sqrt{4 \pi G \rho_s}},
	\label{eq:tc}
\end{equation}
where $C = 0.75$ is a numerical factor and $\sigma_{\mathrm{eff}}$ is the effective cross section~\citep{Yang:2022hkm-SIDM}. For simplicity, we assume a constant cross section in this work. Figure~\ref{fig:cdm} (right) shows the collapse time vs virial mass for the progenitors, where we have taken $\sigma/m=50~{\rm cm^2~g^{-1}}$. About one-third of the halos are expected to collapse within $10~{\rm Gyr}$. Since tidal stripping could speed up the onset of the collapse~\citep{Nishikawa:2019lsc-tid, Sameie:2019zfo-tid,Kahlhoefer:2019oyt-tid,Yang:2021kdf,Zeng:2021ldo-tid}, we expect that more halos would be in the collapse phase after they evolve in the tidal field, and their overall mass would be reduced as well.

\section{Simulation setup}
\label{sec:setup}

In this section, we introduce our simulation setup, including the initial halo density profile, the SIDM cross section, orbital parameters, and the gravitational potential model of the Milky Way. 

\subsection{The initial halo density profile and cross section}
\label{sec:ic}

We choose the CDM progenitor with the earliest infall time among the five halos that have $t_c<10~{\rm Gyr}$ and its density profile is shown in Figure~\ref{fig:cdm} (left, black). For this halo, the fitted NFW parameters are $\rho_s = 7.5\times10^7~ M_\odot~{\rm kpc^{-3}}$ and $r_s = 0.50~{\rm kpc}$. The maximum circular velocity and the associated radius are $V_{\rm max} = 14.8~{\rm km~s^{-1}}$ and $r_{\rm max} = 1.1~{\rm kpc}$, respectively. We use the public code~\texttt{SpherIC}~\citep{Garrison-Kimmel:2013yys} to generate the initial condition and the total halo mass is $M= 3.25 \times 10^8 M_\odot$. The simulation has a particle mass of $32.5~M_\odot$, a total number of $10^7$ particles, and a softening length of $\epsilon = 2~{\rm pc}$. We use the public N-body code \texttt{GADGET-2}~\citep{Springel:2000yr-GADGET2, Springel:2005mi-GADGET2} implemented with an SIDM module from~\cite{Yang:2020iya-SIDM}, which follows the algorithm in~\cite{Robertson:2016xjh} with small modifications. 

As indicated in Figure~\ref{fig:cdm} (right), the halo would collapse within $10~{\rm Gyr}$ for $\sigma/m=50~{\rm cm^2~g^{-1}}$ even it is isolated. In our N-body simulations, we consider three values: $\sigma/m = 30~{\rm cm^2~g^{-1}}$ (SIDM30), $50~{\rm cm^2~g^{-1}}$ (SIDM50), and $100~{\rm cm^2~g^{-1}}$ (SIDM100) to explore a wide range of cross sections. A viable SIDM model should exhibit a velocity-dependent cross section that is large at low velocities while decreasing toward high velocities to evade constraints on massive halos around cluster scales $\lesssim0.1~{\rm cm^2~g^{-1}}$ at $V_{\rm max}\sim1000~{\rm km~s^{-1}}$~\citep{Rocha:2012jg-cons, Peter:2012jh-cons, Harvey:2015hha-cons, Kaplinghat:2015aga-cons,Sagunski:2020spe-cons,Andrade:2020lqq-cons,Ray:2022ydr,Kong:2024zyw}. Nevertheless, for a specific halo, we can use a constant effective cross section to characterize its gravothermal evolution~\citep{Yang:2022hkm-SIDM,Yang:2022zkd,Outmezguine:2022bhq}. In our case, $V_{\rm max}$ decreases from $\sim15~{\rm km~s^{-1}}$ to $7~{\rm km~s^{-1}}$ due to tidal mass loss. Thus the $\sigma/m$ values we consider can be regard as effective cross sections for $V_{\rm max }\sim10~{\rm km~s^{-1}}$ on average, which overall align with SIDM models proposed to explain diverse dark matter distributions in galaxies~\citep[e.g.][]{Valli:2017ktb,Ren:2018jpt,Zavala:2019sjk-tid, Kaplinghat:2019svz,Kahlhoefer:2019oyt-tid,Sameie:2019zfo-tid,Turner:2020vlf,Slone:2021nqd,Yang:2021kdf, Correa:2020qam-tid,Silverman:2022bhs,Correa:2022dey,Gilman:2022ida,Nadler:2023nrd-tid,Fischer:2023lvl,Zhang:2024ggu,ManceraPina:2024ybj,Ragagnin:2024deh,Dutra:2024qac,Roberts:2024uyw}

\subsection{The Milky Way model}
\label{sec:MW}
The Milky Way is modeled as a static potential that contains three main components. 
\begin{itemize}
	\item A spherical NFW halo 
	\begin{equation}
	\Phi_{\text {DM}}(r)=-4 \pi G \rho_s r_s^3 \frac{\ln \left(1+r / r_s\right)}{r},
	\end{equation}
	with $\rho_s = 8.54 \times 10^6  M_\odot ~{\rm kpc^{-3}}$ and $r_s = 19.6~{\rm kpc}$. $G$ is the Newton constant. 
	
	\item A spherical stellar bulge with a Hernquist profile \citep{Hernquist:1990be}:

    \begin{equation}
       \Phi_{\rm b} (r) = -\frac{G M_{\rm b}}{r_{\rm H} + r},
    \end{equation}
    with $M_{\rm b} = 9.23 \times 10^9 M_\odot$ and $r_{\rm H} = 1.3$ kpc. 
	
    \item Two stellar disks and two gas disks with an axisymmetric Miyamoto–Nagai profile~\citep{Miyamoto:1975zz}:
	\begin{equation}
	\Phi_{\rm disc} (R,z) = -\frac{G M_{\rm d}}{\left[ R^2+\left(a_{\rm d}+\sqrt{z^2+b_{\rm d}^2}\right)^2 \right]^{1/2}}.
	\end{equation}
	The parameters for each disk are as follows. Thin stellar disk: $M_{\rm d} = 3.52 \times 10^{10}~M_\odot$, $a_{\rm d} = 2.50~{\rm kpc}$, and $b_{\rm d} = 0.3$ kpc; thick stellar disk: $M_{\rm d} = 1.05 \times 10^{10}~M_\odot$, $a_{\rm d} = 3.02~{\rm kpc}$, and $b_{\rm d} = 0.9~{\rm kpc}$; thin gas disk: $M_{\rm d} = 1.2 \times 10^{9}~M_\odot$, $a_{\rm d} = 1.5~{\rm kpc}$, and $b_{\rm d} = 0.045~{\rm kpc}$; thick gas disk: $M_{\rm d} = 1.1 \times 10^{10}~M_\odot$, $a_{\rm d} = 7.0~{\rm kpc}$, and $b_{\rm d} = 0.085~{\rm kpc}$. 
          
\end{itemize}
These parameters are motivated by the Milky Way mass model in~\cite{McMillan:2016jtx-MW}. Note that the stellar and disk density profiles in~\cite{McMillan:2016jtx-MW} use exponential functions, which are challenging to implement in controlled N-body simulations due to the lack of analytical expressions for their corresponding potentials. Nevertheless, we have verified that the difference in the total potential remains within $2\%$ in the regions with $\sqrt{R^2+z^2}>10~{\rm kpc}$, which are most relevant for our simulated subhalo. Since the host halo is treated as a static potential, we neglect dark matter particle scatterings between the host halo and the subhalo. This approximation is well justified for velocity-dependent SIDM models with $\sigma/m\lesssim1~{\rm cm^2~g^{-1}}$ at $V_{\rm max}\sim200 ~{\rm km~s^{-1}}$~\citep{Nadler:2020ulu}.

\subsection{Orbital parameters}
\label{sec:orbit}

\citet{Bonaca:2020psc} found that the best-fit orbit of GD-1 has a pericenter of $r_{\rm peri}=13.8~\text{kpc}$ and an apocenter of $r_{\rm apo}=22.3~\text{kpc}$, while the orbit of its perturber remains highly uncertain. For our simulation, we adopt an orbit with $r_{\rm peri}=17~\text{kpc}$ and $r_{\rm apo}=142~\text{kpc}$, with the simulated subhalo undergoing five pericenter passages over $10~\text{Gyr}$. Although we do not aim to explicitly model the encounter event, at $t \approx 10~\text{Gyr}$, the simulated subhalo's coordinates are ${\rm R.A.} = 21.5^\circ$ and ${\rm Dec} = -7.9^\circ$, consistent with the inferred position range of the present-day GD-1 perturber~\citep{Bonaca:2020psc}. While this orbit differs from that of the progenitor halo selected from the cosmological merger tree~\citep{Yang:2022mxl-tid-cosmo-CDM}, it remains typical for many subhalos in the simulation. We emphasize that gravothermal collapse is intrinsic to SIDM halos, and the overall properties of our simulated perturber are robust regardless of the specific orbit chosen.

\begin{figure*}[t!]
	\centering
	\includegraphics[scale=0.33]{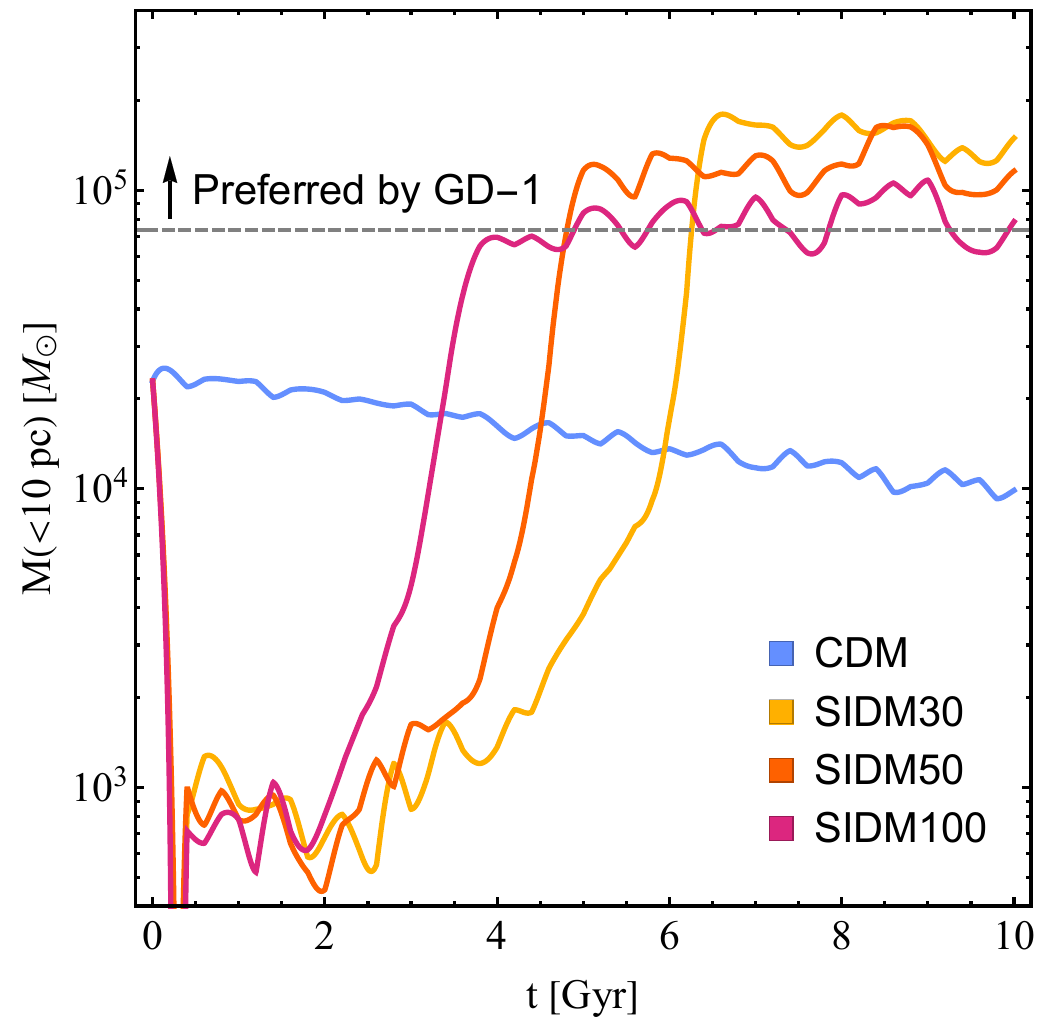}
	\includegraphics[scale=0.33]{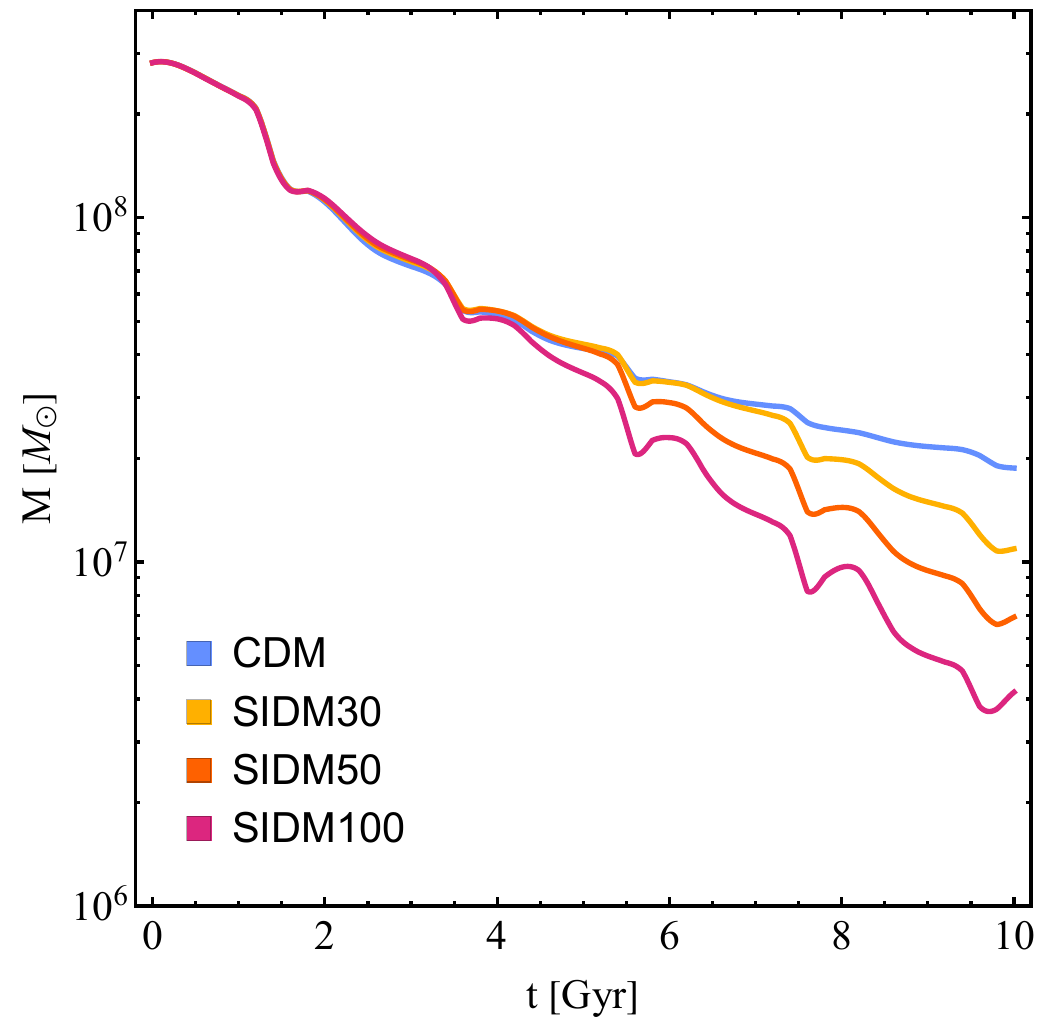}
	\includegraphics[scale=0.22]{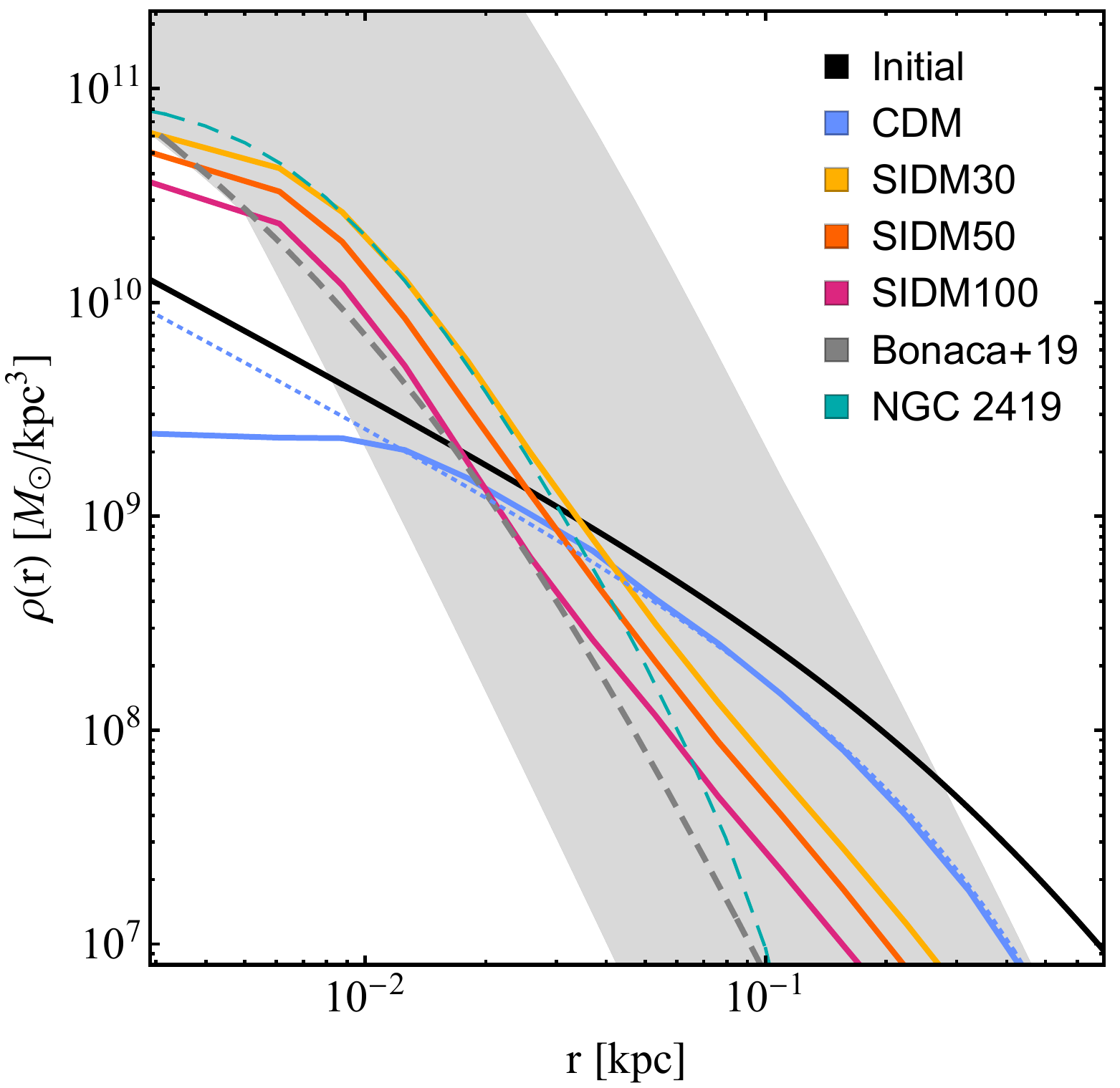}
	\caption{{\bf Left:} evolution of the enclosed mass within $10~{\rm pc}$ for the CDM (blue), SIDM30 (amber), SIDM50 (orange), and SIDM100 (pink) subhalos. The horizontal dashed gray line denotes the inner mass within $10~{\rm pc}$ of the reference Hernquist profile, as shown in Figure~\ref{fig:cdm} (middle). {\bf Middle:} evolution of the bound mass for the simulated CDM and SIDM subhalos. {\bf Right:} corresponding density profiles for the simulated CDM and SIDM subhalos at $t=10~{\rm Gyr}$, along with the initial NFW profile (black). The dotted blue line denotes a reconstructed density profile for the CDM subhalo using an analytical function proposed by~\cite{Errani:2020wgn}. The gray shaded region denotes the viable range for the GD-1 perturber, converted from Figure 6 of~\cite{Bonaca:2018fek-main}, while the dashed gray line represents the reference Hernquist profile. The dashed cyan line represents the density profile of the globular cluster NGC~2419, modeled using the King profile from~\cite{Baumgardt:2009hv}.}
	\label{fig:sidm}
\end{figure*}

\section{Results}
\label{sec:results}

Figure~\ref{fig:sidm} (left) shows the evolution of the enclosed mass within the inner $r=10~{\rm pc}$ for CDM (blue), SIDM30 (amber), SIDM50 (orange), and SIDM100 (pink) subhalos. For CDM, the inner mass decreases monotonically due to tidal stripping. In contrast, for SIDM, the mass initially decreases sharply due to core expansion, followed by an increase as core collapse occurs. By $t \approx 10~{\rm Gyr}$, the inner mass of the SIDM subhalos is an order of magnitude higher than that of the CDM subhalo, aligning well with the reference Hernquist profile (horizontal line). Additionally, the collapse times are $t_c \sim6~{\rm Gyr},~4~{\rm Gyr}$, and $2~{\rm Gyr}$ for SIDM30, SIDM50, and SIDM100, respectively, about a factor of two shorter than those estimated using Equation~\ref{eq:tc}, which is calibrated for isolated halos. In a subhalo, tidal stripping reduces the velocity dispersion of dark matter particles from the intermediate to outer regions as a result of mass loss. Consequently, a negative ``temperature" gradient—a necessary condition for the onset of core collapse—is more easily established compared to an isolated halo~\citep{Sameie:2019zfo-tid}.

In Figure~\ref{fig:sidm} (middle), we show the evolution of the total bound mass for the simulated CDM and SIDM subhalos. Initially, the halo mass is $3.25\times10^8~M_\odot$, and is reduced by an order of magnitude by $t\approx10~{\rm Gyr}$ due to tidal stripping. As expected, the total mass loss is more significant as the cross section increases. For SIDM, the final halo mass ranges from $4\times10^6\textup{--}10^7~M_\odot$, which falls well within the favored mass range of the GD-1 perturber $3\times10^{5}\textup{--}10^8~M_\odot$~\citep{Bonaca:2018fek-main}. 

Figure~\ref{fig:sidm} (right) shows the corresponding density profiles at $t=10~{\rm Gyr}$ for the CDM and SIDM subhalos, along with the initial NFW profile. For comparison, the viable region for the GD-1 perturber (shaded gray), converted from Figure 6 of~\cite{Bonaca:2018fek-main}, and the reference Hernquist profile (dashed gray) are also shown. Compared to CDM, the density profiles of the SIDM subhalos are significantly steeper and overall consistent with the favored Hernquist profiles from~\cite{Bonaca:2018fek-main}. This indicates that dark matter self-interactions can both increase central density and accelerate tidal mass loss in the outer regions. Consequently, an SIDM subhalo can become more compact and dense than its CDM counterpart.   

We note that the CDM subhalo has a small density core near the center. This is due to the resolution limit as $\ell=2.8\epsilon\approx5.6~{\rm pc}$, although the simulated subhalo contains more than $6\times10^{5}$ simulation particles at $t=10~{\rm Gyr}$. We fit the density profile using the analytical function $\rho(r)=\rho_{\rm cut}\exp(-r/r_{\rm cut})r_{\rm cut}/r$ from~\cite{Errani:2020wgn}, which is proposed to model a tidally stripped CDM halo. With $\rho_{\rm cut}\approx1.2\times10^8~M_\odot~{\rm kpc^{-3}}$ and $r_{\rm cut}\approx0.22~{\rm kpc}$, we find a good fit for the region $r\gtrsim10~{\rm pc}$, see Figure~\ref{fig:sidm} (right, dotted blue). The fitted function has a cusp $\rho(r)\propto r^{-1}$ near the center and it provides a correction to the core due to the resolution limit. For the fitted profile, the enclosed mass within $10~{\rm pc}$ is $1.6\times10^4~M_\odot$. However, even with this correction, the CDM subhalo remains insufficiently dense to explain the high density of the GD-1 perturber.

In Figure~\ref{fig:sidm} (right), we also present the density profile of the globular cluster NGC~2419 (cyan), modeled using the King profile from~\cite{Baumgardt:2009hv}. Interestingly, this profile closely resembles the density profile of the SIDM30 subhalo within $30~{\rm pc}$. This similarity is not coincidental, as the formation of globular clusters follows the same mechanism as the collapse of SIDM halos. This suggests that distinguishing between SIDM and globular cluster scenarios in explaining the GD-1 perturbation could be challenging. However, NGC~2419 itself cannot be the GD-1 perturber, as its orbit does not align with the perturbation~\citep{Bonaca:2018fek-main}. If the perturbation is caused by an undetected globular cluster that emits light, it could be identified in future astronomical surveys. 

Furthermore, narrowing down the favored parameter space in the mass-size plane for the perturber would help us distinguish the two scenarios. For instance, if the perturber’s mass is further constrained to the range $10^7\textup{--}10^8~M_\odot$, the SIDM scenario would be favored, as globular clusters typically have masses below a few times $10^{6}~M_\odot$. Another intriguing possibility is that the perturber is an SIDM substructure hosting stars, as we will discuss later. It may have undergone significant tidal stripping, resulting in an ultrafaint dwarf with mass and structural properties similar to those of a massive globular cluster~\citep{DELVE:2019xvr}. Confirming this scenario would require detecting a stellar counterpart at the inferred location of the perturber. Distinguishing between these possibilities will require dedicated observational campaigns and detailed modeling efforts, making this an exciting avenue for future research.

\section{Discussions and conclusion}
\label{sec:diss}

The inner density profiles of our simulated SIDM subhalos ($r\lesssim10~{\rm pc}$) could be underestimated due to numerical issues in N-body simulations when the halo is deeply collapsed~\citep{Zhong:2023yzk,Mace:2024uze,Palubski:2024ibb,Fischer:2024eaz}. Specially, numerical artifacts introduce additional ``energy" that heats the simulated halo, slowing down or even preventing further increases in inner density; see~\cite{Fischer:2024eaz} for discussions about potential causes. As shown in Figure~\ref{fig:sidm} (left), for the SIDM subhalos, the enclosed mass within inner $10~{\rm pc}$ stalls after $t \approx4.5\textup{--}6.5~{\rm Gyr}$, suggesting that they may suffer from the artificial heating effect. To further test this, we conducted an isolated simulation without the tidal field for the same initial NFW profile and $\sigma/m=50~{\rm cm^2~g^{-1}}$. Since the isolated halo experiences no tidal mass loss or heating, its total energy can be computed straightforwardly. See Appendix~\ref{sec:iso} for details on the isolated simulation and comparison with the subhalos.

Indeed, we find that the energy increases when the isolated halo enters the deep collapse phase, corresponding to a Knudsen number of $Kn\approx0.4$ within $10~{\rm pc}$, i.e., the ratio of the mean free path to the gravitational height~\citep[e.g.][]{Balberg:2002ue-CC,Essig:2018pzq-tc}. For the SIDM subhalos, the stalling behavior occurs when their $Kn$ values reach $0.3\textup{--}0.6$. In comparison, the total energy of the simulated SIDM halo in~\cite{Fischer:2024eaz} (their Figure 1) starts to increase when $Kn$ reaches $0.1$. Even at $Kn=0.01$ energy conservation violation is at the $1.5\%$ level, better than our simulation. This is likely because~\cite{Fischer:2024eaz} adopted a more accurate criterion for the gravity computations, while at a higher computational cost. Since the artificial heating effect leads to an underestimation of the inner density profile for a collapsed SIDM halo, our results are conservative in this regard. Nevertheless, it will be important to further improve the SIDM prediction as future measurements of the GD-1 stream could narrow down the viable parameter space of the perturber~\citep{Bonaca:2024dgc}.

When modeling the Milky Way, we used static potentials for both halo and stars, calibrated with present-day measurements. Simulations show that Milky Way-like systems could grow significantly over the last $\sim6~{\rm Gyr}$ due to mergers and accretion~\citep[e.g.][]{Ishchenko:2023,Wang:2024moc}. If these effects were incorporated, our simulated subhalo would experience weaker tidal stripping in the early stages. However, we note that this is degenerate with the orbital parameters; similar results can be achieved by lowering the pericenter if a weaker potential is adopted at early times. In Appendix~\ref{sec:iso}, we will see that even for an isolated halo, the SIDM50 case can still collapse to the viable parameter region. Additionally, encounters between the GD-1 stream and the perturber are likely to have occurred within the last $1~{\rm Gyr}$~\citep{Bonaca:2018fek-main} and hence the growth history of the Milky Way may not directly impact the inference of the perturber's properties.

The subhalo we used to demonstrate the SIDM scenario for the GD-1 perturber has an infall mass of $\approx 3\times 10^8~M_\odot$. Interestingly, this is near the upper limit on the peak mass of subhalos that host currently-observed satellite galaxies in the Milky Way~\citep{Jethwa:2016gra,DES:2019ltu-Nadler2020}. Thus, it remains an open question whether the perturber is a truly dark substructure, devoid of a galaxy. To further investigate detectability, we conducted additional simulations for the CDM and SIDM50 cases with live stellar particles, assuming a Plummer stellar profile with a scale radius of $0.3~{\rm kpc}$ and a total mass of $3.2\times10^4~{M_\odot}$, motivated by hydrodynamical simulations of the Local Group~\citep{Fattahi:2018ioj}. At $t=10~{\rm Gyr}$, the bound stellar masses are $1.8\times10^4~M_\odot$ and $1.3\times10^4~M_\odot$ for the CDM and SIDM50 cases, respectively, with the latter also exhibiting a steeper stellar density profile toward the central regions. These substructures fall into the category of ultrafaint dwarf galaxies and could potentially be detected in the near future through observations, e.g., with the Rubin Observatory~\citep{LSST:2008ijt,LSSTDarkMatterGroup:2019mwo}. 

More work is needed along these lines. For instance, the stellar-halo mass relation becomes increasingly steep in the ultrafaint regime and exhibits significant scatter~\citep{Fattahi:2018ioj}, which must be taken into account. Additionally, since the Rubin Observatory can only detect objects in the southern hemisphere, it would be crucial to assess Rubin's sky coverage in conjunction with the orbital information of the GD-1 perturber from~\cite{Bonaca:2020psc}. We leave these investigations for future work. Furthermore, our scenario should also apply to smaller infall masses below $\sim10^{8}~M_\odot$. Indeed, for SIDM models with large velocity-dependent cross sections, the population of core-collapsing (sub)halos increases as the mass decreases~\citep[e.g.][]{Yang:2022mxl-tid-cosmo-CDM,Nadler:2023nrd-tid}. Thus stellar streams like GD-1 can probe both population and density profile of core-collapsing subhalos even below the mass threshold for galaxy formation.

We used the Hernquist profile for the GD-1 perturber from~\cite{Bonaca:2018fek-main} as a reference to assess the simulated subhalos.  It would be intriguing to take the SIDM subhalo and directly model its encounter with GD-1, incorporating the influence of the Large Magellanic Cloud~\citep[e.g.][]{Erkal:2019,Shipp:2021}. We could use the parametric model~\citep{Yang:2023jwn,Yang:2024uqb, Ando:2024kpk} to generate a population of collapsed SIDM subhalos in Milky Way analogs. To overcome the numerical issues in N-body simulations of core-collapsing halos, we may complement them with the semi-analytical fluid model to better capture the dynamics in the the central regions~\citep[e.g.][]{Balberg:2002ue-CC,Zhong:2023yzk,Gad-Nasr:2023gvf}.

In summary, we have conducted controlled N-body simulations and shown that a core-collapsed SIDM halo could explain the high density of the GD-1 stellar stream perturber. For progenitor halos from the cosmological simulation of a Milky Way analog, the required self-interacting cross section $\sigma/m\gtrsim30~{\rm cm^2~g^{-1}}$ for $\sim10^8~M_\odot$ halos with $V_{\rm max}\sim10~{\rm km~s^{-1}}$. Dark matter self-interactions can both increase inner density and accelerate tidal mass loss in the outer regions, producing a compact and dense perturber to explain the spur and gap features of the GD-1 stream. Our findings demonstrate that stellar streams provide a novel probe into the self-interacting nature of dark matter. We have also outlined future investigations to further improve this promising approach.

\section*{acknowledgments}
We thank Moritz Fischer, Ana Bonaca, and Ting Li for useful discussions and the organizers of the Pollica 2023 SIDM Workshop, where this work was initialized. The work of H.-B.Y. and D.Y. was supported by the John Templeton Foundation under grant ID\#61884 and the U.S. Department of Energy under grant No.~de-sc0008541. This research was supported in part by grant NSF PHY-2309135 to the Kavli Institute for Theoretical Physics (KITP). Computations were performed using the computer clusters and data storage resources of the HPCC at UCR, which were funded by grants from NSF (MRI-2215705, MRI-1429826) and NIH (1S10OD016290-01A1). The opinions expressed in this publication are those of the authors and do not necessarily reflect the views of the funding agencies.

\bibliography{refbib-perturber}

\begin{thebibliography}{}
\expandafter\ifx\csname natexlab\endcsname\relax\def\natexlab#1{#1}\fi
\providecommand{\url}[1]{\href{#1}{#1}}
\providecommand{\dodoi}[1]{doi:~\href{http://doi.org/#1}{\nolinkurl{#1}}}
\providecommand{\doeprint}[1]{\href{http://ascl.net/#1}{\nolinkurl{http://ascl.net/#1}}}
\providecommand{\doarXiv}[1]{\href{https://arxiv.org/abs/#1}{\nolinkurl{https://arxiv.org/abs/#1}}}

\bibitem[{Adhikari {et~al.}(2022)}]{Adhikari:2022sbh}
Adhikari, S., {et~al.} 2022.
\newblock \doarXiv{2207.10638}

\bibitem[{Ando {et~al.}(2024)Ando, Horigome, Nadler, Yang, \&
  Yu}]{Ando:2024kpk}
Ando, S., Horigome, S., Nadler, E.~O., Yang, D., \& Yu, H.-B. 2024.
\newblock \doarXiv{2403.16633}

\bibitem[{Andrade {et~al.}(2021)Andrade, Fuson, Gad-Nasr, Kong, Minor, Roberts,
  \& Kaplinghat}]{Andrade:2020lqq-cons}
Andrade, K.~E., Fuson, J., Gad-Nasr, S., {et~al.} 2021, Mon. Not. Roy. Astron.
  Soc., 510, 54, \dodoi{10.1093/mnras/stab3241}

\bibitem[{Balberg {et~al.}(2002)Balberg, Shapiro, \&
  Inagaki}]{Balberg:2002ue-CC}
Balberg, S., Shapiro, S.~L., \& Inagaki, S. 2002, Astrophys. J., 568, 475,
  \dodoi{10.1086/339038}

\bibitem[{Banik {et~al.}(2021)Banik, Bovy, Bertone, Erkal, \&
  de~Boer}]{Banik:2019cza}
Banik, N., Bovy, J., Bertone, G., Erkal, D., \& de~Boer, T. J.~L. 2021, Mon.
  Not. Roy. Astron. Soc., 502, 2364, \dodoi{10.1093/mnras/stab210}

\bibitem[{Baumgardt {et~al.}(2009)Baumgardt, Cote, Hilker, Rejkuba, Mieske,
  Djorgovski, \& Stetson}]{Baumgardt:2009hv}
Baumgardt, H., Cote, P., Hilker, M., {et~al.} 2009, Mon. Not. Roy. Astron.
  Soc., 396, 2051, \dodoi{10.1111/j.1365-2966.2009.14932.x}

\bibitem[{Bonaca \& Hogg(2018)}]{Bonaca:2018-cons}
Bonaca, A., \& Hogg, D.~W. 2018, The Astrophysical Journal, 867, 101,
  \dodoi{10.3847/1538-4357/aae4da}

\bibitem[{{Bonaca} {et~al.}(2019){Bonaca}, {Hogg}, {Price-Whelan}, \&
  {Conroy}}]{Bonaca:2018fek-main}
{Bonaca}, A., {Hogg}, D.~W., {Price-Whelan}, A.~M., \& {Conroy}, C. 2019, \apj,
  880, 38, \dodoi{10.3847/1538-4357/ab2873}

\bibitem[{Bonaca \& Price-Whelan(2024)}]{Bonaca:2024dgc}
Bonaca, A., \& Price-Whelan, A.~M. 2024.
\newblock \doarXiv{2405.19410}

\bibitem[{Bonaca {et~al.}(2020)Bonaca, Conroy, Hogg, Cargile, Caldwell, Naidu,
  Price-Whelan, Speagle, \& Johnson}]{Bonaca:2020psc}
Bonaca, A., Conroy, C., Hogg, D.~W., {et~al.} 2020, Astrophys. J. Lett., 892,
  L37, \dodoi{10.3847/2041-8213/ab800c}

\bibitem[{Borukhovetskaya {et~al.}(2022)Borukhovetskaya, Navarro, Errani, \&
  Fattahi}]{Borukhovetskaya:2021ahz}
Borukhovetskaya, A., Navarro, J.~F., Errani, R., \& Fattahi, A. 2022, Mon. Not.
  Roy. Astron. Soc., 512, 5247, \dodoi{10.1093/mnras/stac653}

\bibitem[{Bovy {et~al.}(2016)Bovy, Bahmanyar, Fritz, \&
  Kallivayalil}]{Bovy:2016chl-cons}
Bovy, J., Bahmanyar, A., Fritz, T.~K., \& Kallivayalil, N. 2016, Astrophys. J.,
  833, 31, \dodoi{10.3847/1538-4357/833/1/31}

\bibitem[{Carlberg \& Grillmair(2013)}]{Carlberg:2013gxa}
Carlberg, R.~G., \& Grillmair, C.~J. 2013, Astrophys. J., 768, 171,
  \dodoi{10.1088/0004-637X/768/2/171}

\bibitem[{Correa(2021)}]{Correa:2020qam-tid}
Correa, C.~A. 2021, Mon. Not. Roy. Astron. Soc., 503, 920,
  \dodoi{10.1093/mnras/stab506}

\bibitem[{Correa {et~al.}(2022)Correa, Schaller, Ploeckinger, Anau~Montel,
  Weniger, \& Ando}]{Correa:2022dey}
Correa, C.~A., Schaller, M., Ploeckinger, S., {et~al.} 2022, Mon. Not. Roy.
  Astron. Soc., 517, 3045, \dodoi{10.1093/mnras/stac2830}

\bibitem[{de~Boer {et~al.}(2018)de~Boer, Belokurov, Koposov, Ferrarese, Erkal,
  Côté, \& Navarro}]{deBoer:2018-gap}
de~Boer, T. J.~L., Belokurov, V., Koposov, S.~E., {et~al.} 2018, Monthly
  Notices of the Royal Astronomical Society, 477, 1893–1902,
  \dodoi{10.1093/mnras/sty677}

\bibitem[{{de Boer} {et~al.}(2020){de Boer}, {Erkal}, \&
  {Gieles}}]{deBoer:2020}
{de Boer}, T.~J.~L., {Erkal}, D., \& {Gieles}, M. 2020, \mnras, 494, 5315,
  \dodoi{10.1093/mnras/staa917}

\bibitem[{Doke \& Hattori(2022)}]{Doke:2022jro-gaia3}
Doke, Y., \& Hattori, K. 2022, Astrophys. J., 941, 129,
  \dodoi{10.3847/1538-4357/aca090}

\bibitem[{Drlica-Wagner {et~al.}(2019)}]{LSSTDarkMatterGroup:2019mwo}
Drlica-Wagner, A., {et~al.} 2019.
\newblock \doarXiv{1902.01055}

\bibitem[{Dutra {et~al.}(2024)Dutra, Natarajan, \& Gilman}]{Dutra:2024qac}
Dutra, I., Natarajan, P., \& Gilman, D. 2024.
\newblock \doarXiv{2406.17024}

\bibitem[{{Erkal} {et~al.}(2019){Erkal}, {Belokurov}, {Laporte}, {Koposov},
  {Li}, {Grillmair}, {Kallivayalil}, {Price-Whelan}, {Evans}, {Hawkins},
  {Hendel}, {Mateu}, {Navarro}, {del Pino}, {Slater}, {Sohn}, \& {Orphan Aspen
  Treasury Collaboration}}]{Erkal:2019}
{Erkal}, D., {Belokurov}, V., {Laporte}, C.~F.~P., {et~al.} 2019, \mnras, 487,
  2685, \dodoi{10.1093/mnras/stz1371}

\bibitem[{Errani \& Navarro(2021)}]{Errani:2020wgn}
Errani, R., \& Navarro, J.~F. 2021, Mon. Not. Roy. Astron. Soc., 505, 18,
  \dodoi{10.1093/mnras/stab1215}

\bibitem[{Essig {et~al.}(2019)Essig, Mcdermott, Yu, \&
  Zhong}]{Essig:2018pzq-tc}
Essig, R., Mcdermott, S.~D., Yu, H.-B., \& Zhong, Y.-M. 2019, Phys. Rev. Lett.,
  123, 121102, \dodoi{10.1103/PhysRevLett.123.121102}

\bibitem[{Fattahi {et~al.}(2018)Fattahi, Navarro, Frenk, Oman, Sawala, \&
  Schaller}]{Fattahi:2018ioj}
Fattahi, A., Navarro, J., Frenk, C., {et~al.} 2018, Mon. Not. Roy. Astron.
  Soc., 476, 3816, \dodoi{10.1093/mnras/sty408}

\bibitem[{Feng {et~al.}(2021)Feng, Yu, \& Zhong}]{Feng:2020kxv}
Feng, W.-X., Yu, H.-B., \& Zhong, Y.-M. 2021, Astrophys. J. Lett., 914, L26,
  \dodoi{10.3847/2041-8213/ac04b0}

\bibitem[{Fischer {et~al.}(2024{\natexlab{a}})Fischer, Dolag, \&
  Yu}]{Fischer:2024eaz}
Fischer, M.~S., Dolag, K., \& Yu, H.-B. 2024{\natexlab{a}}, Astron. Astrophys.,
  689, A300, \dodoi{10.1051/0004-6361/202449849}

\bibitem[{Fischer {et~al.}(2024{\natexlab{b}})Fischer, Kasselmann, Br\"uggen,
  Dolag, Kahlhoefer, Ragagnin, Robertson, \& Schmidt-Hoberg}]{Fischer:2023lvl}
Fischer, M.~S., Kasselmann, L., Br\"uggen, M., {et~al.} 2024{\natexlab{b}},
  Mon. Not. Roy. Astron. Soc., 529, 2327, \dodoi{10.1093/mnras/stae699}

\bibitem[{Gad-Nasr {et~al.}(2024)Gad-Nasr, Boddy, Kaplinghat, Outmezguine, \&
  Sagunski}]{Gad-Nasr:2023gvf}
Gad-Nasr, S., Boddy, K.~K., Kaplinghat, M., Outmezguine, N.~J., \& Sagunski, L.
  2024, JCAP, 05, 131, \dodoi{10.1088/1475-7516/2024/05/131}

\bibitem[{Garrison-Kimmel {et~al.}(2013)Garrison-Kimmel, Rocha, Boylan-Kolchin,
  Bullock, \& Lally}]{Garrison-Kimmel:2013yys}
Garrison-Kimmel, S., Rocha, M., Boylan-Kolchin, M., Bullock, J., \& Lally, J.
  2013, Mon. Not. Roy. Astron. Soc., 433, 3539, \dodoi{10.1093/mnras/stt984}

\bibitem[{Gilman {et~al.}(2023)Gilman, Zhong, \& Bovy}]{Gilman:2022ida}
Gilman, D., Zhong, Y.-M., \& Bovy, J. 2023, Phys. Rev. D, 107, 103008,
  \dodoi{10.1103/PhysRevD.107.103008}

\bibitem[{Grillmair \& Dionatos(2006)}]{Grillmair:2006bd}
Grillmair, C.~J., \& Dionatos, O. 2006, Astrophys. J. Lett., 643, L17,
  \dodoi{10.1086/505111}

\bibitem[{Harvey {et~al.}(2015)Harvey, Massey, Kitching, Taylor, \&
  Tittley}]{Harvey:2015hha-cons}
Harvey, D., Massey, R., Kitching, T., Taylor, A., \& Tittley, E. 2015, Science,
  347, 1462, \dodoi{10.1126/science.1261381}

\bibitem[{Hernquist(1990)}]{Hernquist:1990be}
Hernquist, L. 1990, Astrophys. J., 356, 359, \dodoi{10.1086/168845}

\bibitem[{{Ishchenko} {et~al.}(2023){Ishchenko}, {Sobolenko}, {Berczik},
  {Khoperskov}, {Omarov}, {Sobodar}, \& {Makukov}}]{Ishchenko:2023}
{Ishchenko}, M., {Sobolenko}, M., {Berczik}, P., {et~al.} 2023, \aap, 673,
  A152, \dodoi{10.1051/0004-6361/202245117}

\bibitem[{Ivezi\'c {et~al.}(2019)}]{LSST:2008ijt}
Ivezi\'c, v., {et~al.} 2019, Astrophys. J., 873, 111,
  \dodoi{10.3847/1538-4357/ab042c}

\bibitem[{Jethwa {et~al.}(2018)Jethwa, Erkal, \& Belokurov}]{Jethwa:2016gra}
Jethwa, P., Erkal, D., \& Belokurov, V. 2018, Mon. Not. Roy. Astron. Soc., 473,
  2060, \dodoi{10.1093/mnras/stx2330}

\bibitem[{Kahlhoefer {et~al.}(2019)Kahlhoefer, Kaplinghat, Slatyer, \&
  Wu}]{Kahlhoefer:2019oyt-tid}
Kahlhoefer, F., Kaplinghat, M., Slatyer, T.~R., \& Wu, C.-L. 2019, JCAP, 12,
  010, \dodoi{10.1088/1475-7516/2019/12/010}

\bibitem[{Kaplinghat {et~al.}(2016)Kaplinghat, Tulin, \&
  Yu}]{Kaplinghat:2015aga-cons}
Kaplinghat, M., Tulin, S., \& Yu, H.-B. 2016, Phys. Rev. Lett., 116, 041302,
  \dodoi{10.1103/PhysRevLett.116.041302}

\bibitem[{Kaplinghat {et~al.}(2019)Kaplinghat, Valli, \&
  Yu}]{Kaplinghat:2019svz}
Kaplinghat, M., Valli, M., \& Yu, H.-B. 2019, Mon. Not. Roy. Astron. Soc., 490,
  231, \dodoi{10.1093/mnras/stz2511}

\bibitem[{Koda \& Shapiro(2011)}]{Koda:2011yb-CC}
Koda, J., \& Shapiro, P.~R. 2011, Mon. Not. Roy. Astron. Soc., 415, 1125,
  \dodoi{10.1111/j.1365-2966.2011.18684.x}

\bibitem[{Kong {et~al.}(2024)Kong, Yang, \& Yu}]{Kong:2024zyw}
Kong, D., Yang, D., \& Yu, H.-B. 2024, Astrophys. J. Lett., 965, L19,
  \dodoi{10.3847/2041-8213/ad394b}

\bibitem[{Koposov {et~al.}(2010)Koposov, Rix, \& Hogg}]{Koposov:2009hn-cons}
Koposov, S.~E., Rix, H.-W., \& Hogg, D.~W. 2010, Astrophys. J., 712, 260,
  \dodoi{10.1088/0004-637X/712/1/260}

\bibitem[{Li {et~al.}(2022)}]{S5:2021ooa}
Li, T.~S., {et~al.} 2022, Astrophys. J., 928, 30,
  \dodoi{10.3847/1538-4357/ac46d3}

\bibitem[{Mace {et~al.}(2024)Mace, Zeng, Peter, Du, Yang, Benson, \&
  Vogelsberger}]{Mace:2024uze}
Mace, C., Zeng, Z.~C., Peter, A. H.~G., {et~al.} 2024.
\newblock \doarXiv{2402.01604}

\bibitem[{Malhan \& Ibata(2019)}]{Malhan:2019-cons}
Malhan, K., \& Ibata, R.~A. 2019, Monthly Notices of the Royal Astronomical
  Society, 486, 2995–3005, \dodoi{10.1093/mnras/stz1035}

\bibitem[{{Malhan} {et~al.}(2022){Malhan}, {Valluri}, {Freese}, \&
  {Ibata}}]{Malhan:2022}
{Malhan}, K., {Valluri}, M., {Freese}, K., \& {Ibata}, R.~A. 2022, \apjl, 941,
  L38, \dodoi{10.3847/2041-8213/aca6e5}

\bibitem[{Mancera Pi\~na {et~al.}(2024)Mancera Pi\~na, Golini, Trujillo, \&
  Montes}]{ManceraPina:2024ybj}
Mancera Pi\~na, P.~E., Golini, G., Trujillo, I., \& Montes, M. 2024.
\newblock \doarXiv{2404.06537}

\bibitem[{Mao {et~al.}(2015)Mao, Williamson, \& Wechsler}]{Mao:2015yua}
Mao, Y.-Y., Williamson, M., \& Wechsler, R.~H. 2015, Astrophys. J., 810, 21,
  \dodoi{10.1088/0004-637X/810/1/21}

\bibitem[{Mau {et~al.}(2020)}]{DELVE:2019xvr}
Mau, S., {et~al.} 2020, Astrophys. J., 890, 136,
  \dodoi{10.3847/1538-4357/ab6c67}

\bibitem[{McMillan(2016)}]{McMillan:2016jtx-MW}
McMillan, P.~J. 2016, Mon. Not. Roy. Astron. Soc., 465, 76,
  \dodoi{10.1093/mnras/stw2759}

\bibitem[{Minor {et~al.}(2021)Minor, Gad-Nasr, Kaplinghat, \&
  Vegetti}]{Minor:2020hic}
Minor, Q.~E., Gad-Nasr, S., Kaplinghat, M., \& Vegetti, S. 2021, Mon. Not. Roy.
  Astron. Soc., 507, 1662, \dodoi{10.1093/mnras/stab2247}

\bibitem[{Miyamoto \& Nagai(1975)}]{Miyamoto:1975zz}
Miyamoto, M., \& Nagai, R. 1975, Publ. Astron. Soc. Jap., 27, 533

\bibitem[{Nadler {et~al.}(2020{\natexlab{a}})Nadler, Banerjee, Adhikari, Mao,
  \& Wechsler}]{Nadler:2020ulu}
Nadler, E.~O., Banerjee, A., Adhikari, S., Mao, Y.-Y., \& Wechsler, R.~H.
  2020{\natexlab{a}}, Astrophys. J., 896, 112, \dodoi{10.3847/1538-4357/ab94b0}

\bibitem[{Nadler {et~al.}(2023)Nadler, Yang, \& Yu}]{Nadler:2023nrd-tid}
Nadler, E.~O., Yang, D., \& Yu, H.-B. 2023, Astrophys. J. Lett., 958, L39,
  \dodoi{10.3847/2041-8213/ad0e09}

\bibitem[{Nadler {et~al.}(2020{\natexlab{b}})}]{DES:2019ltu-Nadler2020}
Nadler, E.~O., {et~al.} 2020{\natexlab{b}}, Astrophys. J., 893, 48,
  \dodoi{10.3847/1538-4357/ab846a}

\bibitem[{Nishikawa {et~al.}(2020)Nishikawa, Boddy, \&
  Kaplinghat}]{Nishikawa:2019lsc-tid}
Nishikawa, H., Boddy, K.~K., \& Kaplinghat, M. 2020, Phys. Rev. D, 101, 063009,
  \dodoi{10.1103/PhysRevD.101.063009}

\bibitem[{Outmezguine {et~al.}(2023)Outmezguine, Boddy, Gad-Nasr, Kaplinghat,
  \& Sagunski}]{Outmezguine:2022bhq}
Outmezguine, N.~J., Boddy, K.~K., Gad-Nasr, S., Kaplinghat, M., \& Sagunski, L.
  2023, Mon. Not. Roy. Astron. Soc., 523, 4786, \dodoi{10.1093/mnras/stad1705}

\bibitem[{Palubski {et~al.}(2024)Palubski, Slone, Kaplinghat, Lisanti, \&
  Jiang}]{Palubski:2024ibb}
Palubski, I., Slone, O., Kaplinghat, M., Lisanti, M., \& Jiang, F. 2024.
\newblock \doarXiv{2402.12452}

\bibitem[{Peter {et~al.}(2013)Peter, Rocha, Bullock, \&
  Kaplinghat}]{Peter:2012jh-cons}
Peter, A. H.~G., Rocha, M., Bullock, J.~S., \& Kaplinghat, M. 2013, Mon. Not.
  Roy. Astron. Soc., 430, 105, \dodoi{10.1093/mnras/sts535}

\bibitem[{Pollack {et~al.}(2015)Pollack, Spergel, \&
  Steinhardt}]{Pollack:2014rja-tc}
Pollack, J., Spergel, D.~N., \& Steinhardt, P.~J. 2015, Astrophys. J., 804,
  131, \dodoi{10.1088/0004-637X/804/2/131}

\bibitem[{Price-Whelan \& Bonaca(2018)}]{Price:2018-gap}
Price-Whelan, A.~M., \& Bonaca, A. 2018, The Astrophysical Journal Letters,
  863, L20, \dodoi{10.3847/2041-8213/aad7b5}

\bibitem[{Ragagnin {et~al.}(2024)Ragagnin, Meneghetti, Calura, Despali, Dolag,
  Fischer, Giocoli, \& Moscardini}]{Ragagnin:2024deh}
Ragagnin, A., Meneghetti, M., Calura, F., {et~al.} 2024, Astron. Astrophys.,
  687, A270, \dodoi{10.1051/0004-6361/202449872}

\bibitem[{Ray {et~al.}(2022)Ray, Sarkar, \& Shaw}]{Ray:2022ydr}
Ray, T.~S., Sarkar, S., \& Shaw, A.~K. 2022, JCAP, 09, 011,
  \dodoi{10.1088/1475-7516/2022/09/011}

\bibitem[{Ren {et~al.}(2019)Ren, Kwa, Kaplinghat, \& Yu}]{Ren:2018jpt}
Ren, T., Kwa, A., Kaplinghat, M., \& Yu, H.-B. 2019, Phys. Rev. X, 9, 031020,
  \dodoi{10.1103/PhysRevX.9.031020}

\bibitem[{Roberts {et~al.}(2024)Roberts, Kaplinghat, Valli, \&
  Yu}]{Roberts:2024uyw}
Roberts, M.~G., Kaplinghat, M., Valli, M., \& Yu, H.-B. 2024.
\newblock \doarXiv{2407.15005}

\bibitem[{Robertson {et~al.}(2017)Robertson, Massey, \&
  Eke}]{Robertson:2016xjh}
Robertson, A., Massey, R., \& Eke, V. 2017, Mon. Not. Roy. Astron. Soc., 465,
  569, \dodoi{10.1093/mnras/stw2670}

\bibitem[{Rocha {et~al.}(2013)Rocha, Peter, Bullock, Kaplinghat,
  Garrison-Kimmel, Onorbe, \& Moustakas}]{Rocha:2012jg-cons}
Rocha, M., Peter, A. H.~G., Bullock, J.~S., {et~al.} 2013, Mon. Not. Roy.
  Astron. Soc., 430, 81, \dodoi{10.1093/mnras/sts514}

\bibitem[{Sagunski {et~al.}(2021)Sagunski, Gad-Nasr, Colquhoun, Robertson, \&
  Tulin}]{Sagunski:2020spe-cons}
Sagunski, L., Gad-Nasr, S., Colquhoun, B., Robertson, A., \& Tulin, S. 2021,
  JCAP, 01, 024, \dodoi{10.1088/1475-7516/2021/01/024}

\bibitem[{Sameie {et~al.}(2020)Sameie, Yu, Sales, Vogelsberger, \&
  Zavala}]{Sameie:2019zfo-tid}
Sameie, O., Yu, H.-B., Sales, L.~V., Vogelsberger, M., \& Zavala, J. 2020,
  Phys. Rev. Lett., 124, 141102, \dodoi{10.1103/PhysRevLett.124.141102}

\bibitem[{Shipp {et~al.}(2018)}]{DES:2018imd}
Shipp, N., {et~al.} 2018, Astrophys. J., 862, 114,
  \dodoi{10.3847/1538-4357/aacdab}

\bibitem[{{Shipp} {et~al.}(2021){Shipp}, {Erkal}, {Drlica-Wagner}, {Li},
  {Pace}, {Koposov}, {Cullinane}, {Da Costa}, {Ji}, {Kuehn}, {Lewis}, {Mackey},
  {Simpson}, {Wan}, {Zucker}, {Bland-Hawthorn}, {Ferguson}, {Lilleengen}, \&
  {Lilleengen}}]{Shipp:2021}
{Shipp}, N., {Erkal}, D., {Drlica-Wagner}, A., {et~al.} 2021, \apj, 923, 149,
  \dodoi{10.3847/1538-4357/ac2e93}

\bibitem[{Silverman {et~al.}(2022)Silverman, Bullock, Kaplinghat, Robles, \&
  Valli}]{Silverman:2022bhs}
Silverman, M., Bullock, J.~S., Kaplinghat, M., Robles, V.~H., \& Valli, M.
  2022, Mon. Not. Roy. Astron. Soc., 518, 2418, \dodoi{10.1093/mnras/stac3232}

\bibitem[{Slone {et~al.}(2023)Slone, Jiang, Lisanti, \&
  Kaplinghat}]{Slone:2021nqd}
Slone, O., Jiang, F., Lisanti, M., \& Kaplinghat, M. 2023, Phys. Rev. D, 107,
  043014, \dodoi{10.1103/PhysRevD.107.043014}

\bibitem[{Springel(2005)}]{Springel:2005mi-GADGET2}
Springel, V. 2005, Mon. Not. Roy. Astron. Soc., 364, 1105,
  \dodoi{10.1111/j.1365-2966.2005.09655.x}

\bibitem[{Springel {et~al.}(2001)Springel, Yoshida, \&
  White}]{Springel:2000yr-GADGET2}
Springel, V., Yoshida, N., \& White, S. D.~M. 2001, New Astron., 6, 79,
  \dodoi{10.1016/S1384-1076(01)00042-2}

\bibitem[{Tulin \& Yu(2018)}]{Tulin:2017ara-core}
Tulin, S., \& Yu, H.-B. 2018, Phys. Rept., 730, 1,
  \dodoi{10.1016/j.physrep.2017.11.004}

\bibitem[{Turner {et~al.}(2021)Turner, Lovell, Zavala, \&
  Vogelsberger}]{Turner:2020vlf}
Turner, H.~C., Lovell, M.~R., Zavala, J., \& Vogelsberger, M. 2021, Mon. Not.
  Roy. Astron. Soc., 505, 5327, \dodoi{10.1093/mnras/stab1725}

\bibitem[{Valli \& Yu(2018)}]{Valli:2017ktb}
Valli, M., \& Yu, H.-B. 2018, Nature Astron., 2, 907,
  \dodoi{10.1038/s41550-018-0560-7}

\bibitem[{Vegetti {et~al.}(2010)Vegetti, Koopmans, Bolton, Treu, \&
  Gavazzi}]{Vegetti:2009cz}
Vegetti, S., Koopmans, L. V.~E., Bolton, A., Treu, T., \& Gavazzi, R. 2010,
  Mon. Not. Roy. Astron. Soc., 408, 1969,
  \dodoi{10.1111/j.1365-2966.2010.16865.x}

\bibitem[{Wang {et~al.}(2024)Wang, Mansfield, Nadler, Darragh-Ford, Wechsler,
  Yang, \& Yu}]{Wang:2024moc}
Wang, Y., Mansfield, P., Nadler, E.~O., {et~al.} 2024.
\newblock \doarXiv{2408.01487}

\bibitem[{Yang {et~al.}(2023{\natexlab{a}})Yang, Nadler, \&
  Yu}]{Yang:2022mxl-tid-cosmo-CDM}
Yang, D., Nadler, E.~O., \& Yu, H.-B. 2023{\natexlab{a}}, Astrophys. J., 949,
  67, \dodoi{10.3847/1538-4357/acc73e}

\bibitem[{Yang {et~al.}(2024{\natexlab{a}})Yang, Nadler, \& Yu}]{Yang:2024uqb}
---. 2024{\natexlab{a}}.
\newblock \doarXiv{2406.10753}

\bibitem[{Yang {et~al.}(2024{\natexlab{b}})Yang, Nadler, Yu, \&
  Zhong}]{Yang:2023jwn}
Yang, D., Nadler, E.~O., Yu, H.-B., \& Zhong, Y.-M. 2024{\natexlab{b}}, JCAP,
  02, 032, \dodoi{10.1088/1475-7516/2024/02/032}

\bibitem[{Yang \& Yu(2021)}]{Yang:2021kdf}
Yang, D., \& Yu, H.-B. 2021, Phys. Rev. D, 104, 103031,
  \dodoi{10.1103/PhysRevD.104.103031}

\bibitem[{Yang \& Yu(2022)}]{Yang:2022hkm-SIDM}
---. 2022, JCAP, 09, 077, \dodoi{10.1088/1475-7516/2022/09/077}

\bibitem[{Yang {et~al.}(2020)Yang, Yu, \& An}]{Yang:2020iya-SIDM}
Yang, D., Yu, H.-B., \& An, H. 2020, Phys. Rev. Lett., 125, 111105,
  \dodoi{10.1103/PhysRevLett.125.111105}

\bibitem[{Yang {et~al.}(2023{\natexlab{b}})Yang, Du, Zeng, Benson, Jiang,
  Nadler, \& Peter}]{Yang:2022zkd}
Yang, S., Du, X., Zeng, Z.~C., {et~al.} 2023{\natexlab{b}}, Astrophys. J., 946,
  47, \dodoi{10.3847/1538-4357/acbd49}

\bibitem[{Zavala {et~al.}(2019)Zavala, Lovell, Vogelsberger, \&
  Burger}]{Zavala:2019sjk-tid}
Zavala, J., Lovell, M.~R., Vogelsberger, M., \& Burger, J.~D. 2019, Phys. Rev.
  D, 100, 063007, \dodoi{10.1103/PhysRevD.100.063007}

\bibitem[{Zeng {et~al.}(2022)Zeng, Peter, Du, Benson, Kim, Jiang, Cyr-Racine,
  \& Vogelsberger}]{Zeng:2021ldo-tid}
Zeng, Z.~C., Peter, A. H.~G., Du, X., {et~al.} 2022, Mon. Not. Roy. Astron.
  Soc., 513, 4845, \dodoi{10.1093/mnras/stac1094}

\bibitem[{Zhang {et~al.}(2024)Zhang, Yu, Yang, \& An}]{Zhang:2024ggu}
Zhang, X., Yu, H.-B., Yang, D., \& An, H. 2024, Astrophys. J. Lett., 968, L13,
  \dodoi{10.3847/2041-8213/ad50cd}

\bibitem[{Zhong {et~al.}(2023)Zhong, Yang, \& Yu}]{Zhong:2023yzk}
Zhong, Y.-M., Yang, D., \& Yu, H.-B. 2023, Mon. Not. Roy. Astron. Soc., 526,
  758, \dodoi{10.1093/mnras/stad2765}

\end{thebibliography}
\bibliographystyle{aasjournal}

\clearpage
\appendix

\section{Simulating halos in the deep collapse phase}
\label{sec:iso}

\begin{figure*}[h]
	\centering
	\includegraphics[scale=0.33]{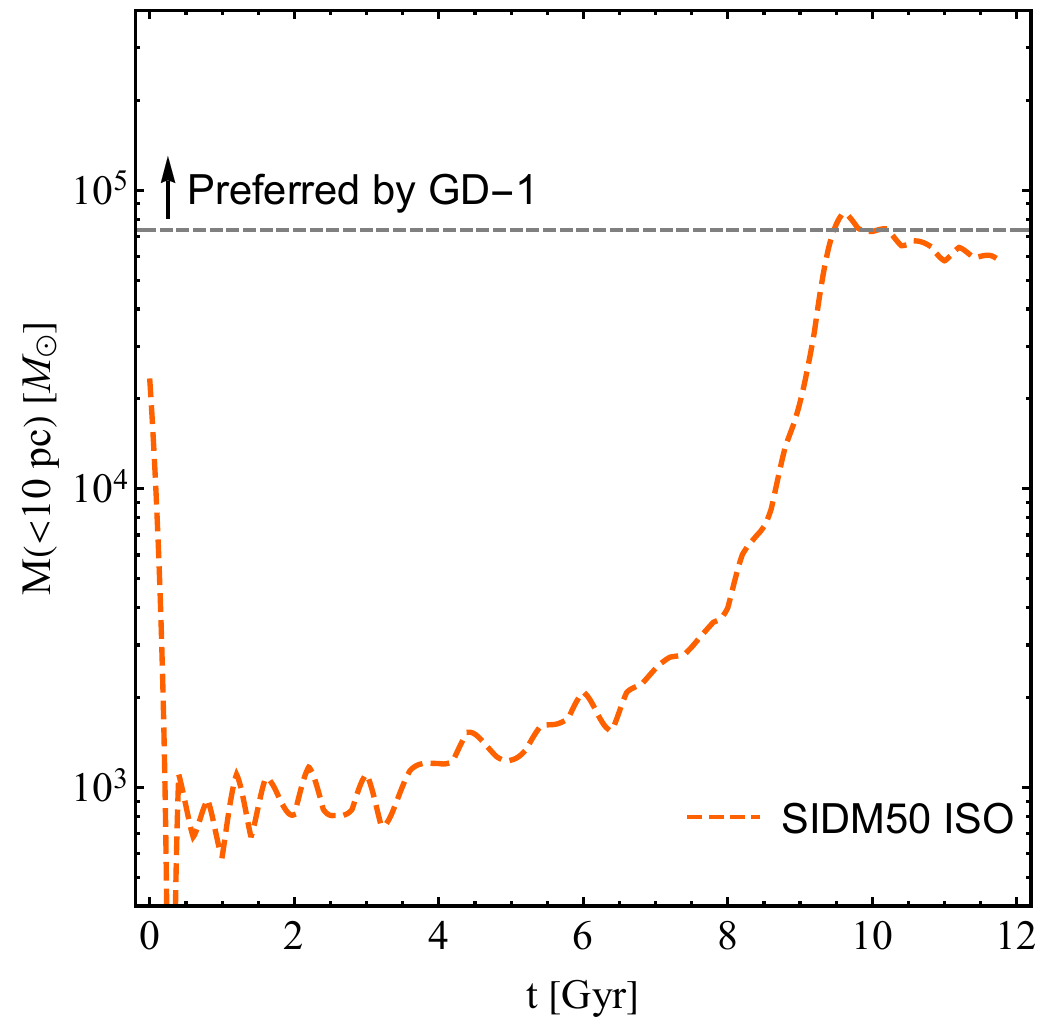}~~
	\includegraphics[scale=0.34]{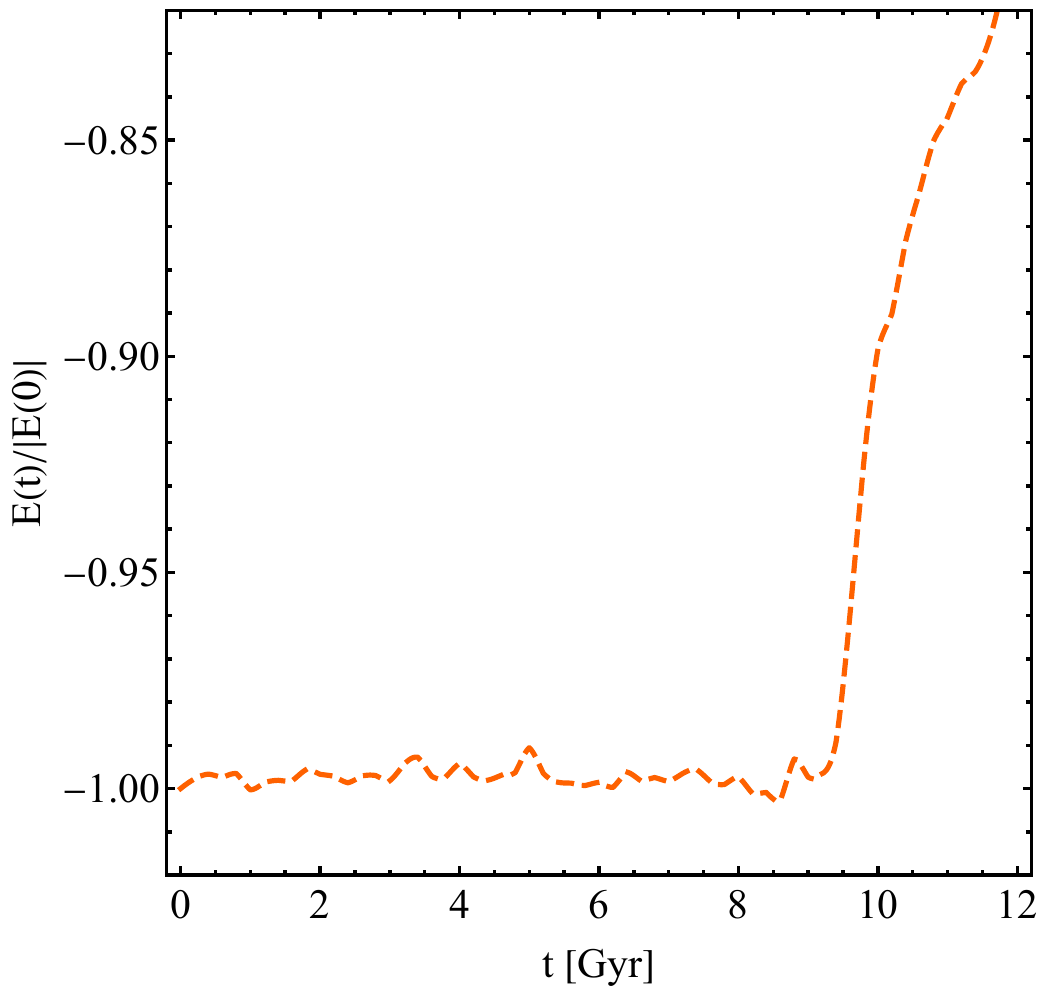}~~
        \includegraphics[scale=0.33]{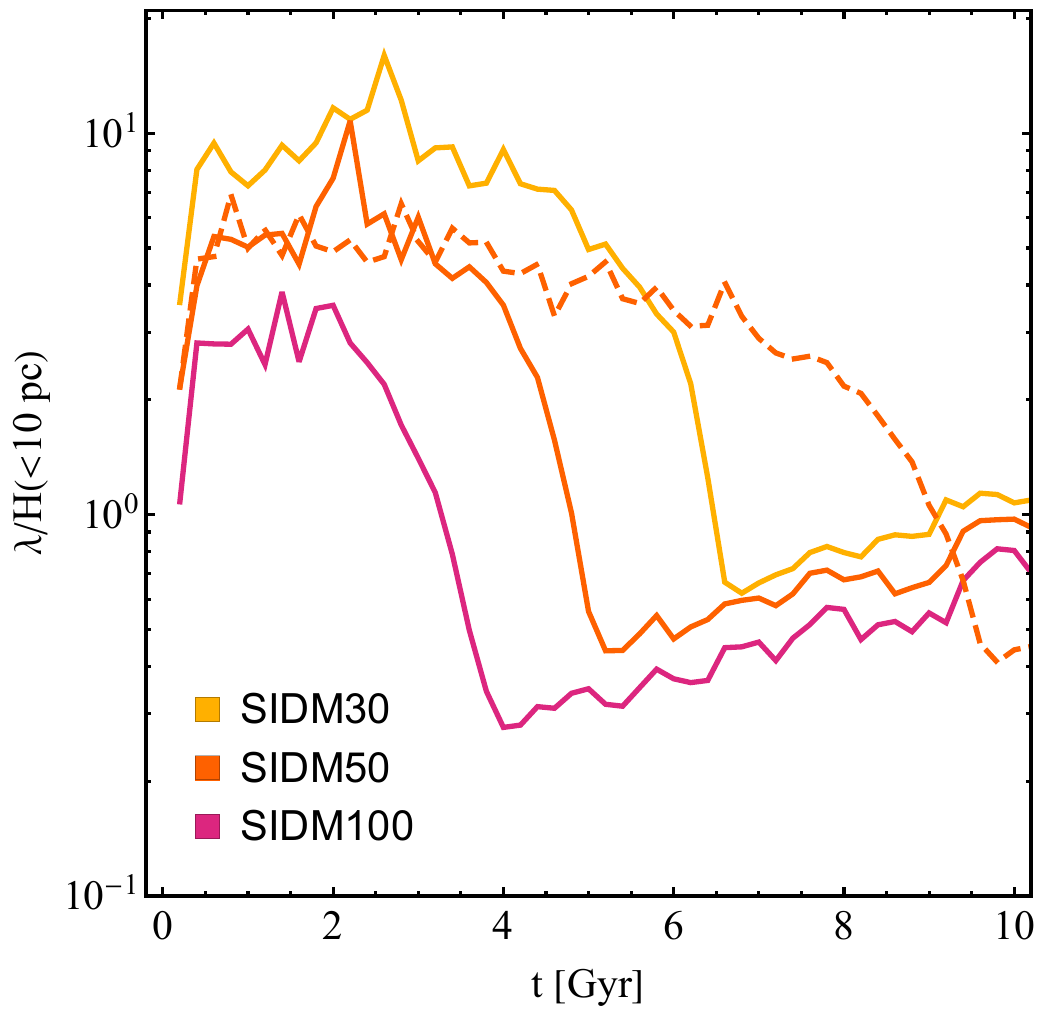}
	\caption{{\bf Left:} evolution of the enclosed mass within $10~{\rm pc}$ for the SIDM50 isolated halo without evolving in the tidal field (dashed orange). The horizontal line denotes the Hernquist profile as in Figure~\ref{fig:sidm} (left). {\bf Middle:} evolution of the total energy normalized to its initial absolute value for the SIDM50 isolated halo. {\bf Right:} evolution of the ratio $\lambda/H$ within inner $10~{\rm pc}$ for the SIDM50 isolated halo (dashed orange), along with the SIDM30 (solid amber), SIDM50 (solid orange), and SIDM100 (solid pink) subhalos.}
	\label{fig:iso}
\end{figure*}

N-body simulations of a core-collapsing SIDM halo are challenging~\citep{Zhong:2023yzk,Mace:2024uze,Palubski:2024ibb,Fischer:2024eaz}. In particular, when a halo enters the phase of deep collapse, energy conservation can be violated in the simulation, resulting in an increase in total energy due to numerical artifacts; see~\cite{Fischer:2024eaz} for discussions on potential causes. This ``heating" effect slows down the further collapse of the simulated halo and the increase of its inner density. As shown in Figure~\ref{fig:sidm} (left), the enclosed mass of the SIDM subhalos stalls at late stages, indicating that they may be affected by these numerical artifacts. To assess the condition of energy conservation, we simulate an {\it isolated} halo with the same initial NFW profile and $\sigma/m=50~{\rm cm^2~g^{-1}}$, without evolving it in the tidal field. For an isolated halo, it is straightforward to evaluate its total energy over time, allowing us to test our simulation setup.

In Figure~\ref{fig:iso} (left), we illustrate the evolution of the enclosed mass within the inner $10~{\rm pc}$ of the isolated SIDM50 halo. The overall behavior is similar to that of the SIDM50 subhalo presented in Figure~\ref{fig:sidm} (left), but the collapse timescale for the isolated halo is approximately a factor of two longer due to the absence of tidal acceleration. After the inner mass reaches its peak at $t\approx9.5~{\rm Gyr}$, it stops increasing and instead experiences a slight decrease. Figure~\ref{fig:iso} (middle) shows the evolution of the total energy of the isolated halo, normalized to its initial absolute value. We observe that the total energy deviates significantly from its initial value for $t > 9.5~{\rm Gyr}$ due to the artificial heating effect.

To further clarify this issue, we compute the Knudsen number $Kn \equiv \lambda/H$, where $\lambda$ is the mean free path and $H$ is the gravitational scale height, i.e.,
\begin{equation}
\lambda=\frac{1}{\rho\sigma/m},~H= \sqrt{\frac{ \sigma^2_{\rm v}}{4\pi G \rho}},
\end{equation}
where $G$ is Newton's constant and $\sigma_{\rm v}$ is the 1D velocity dispersion of dark matter particles. Figure~\ref{fig:iso} (right) shows the evolution of the Knudsen number averaged over inner $10~{\rm pc}$ for the SIDM50 isolated halo (dashed orange), and the SIDM30 (solid amber), SIDM50 (solid orange), and SIDM100 (solid pink) subhalos. At $t\approx9.5~{\rm Gyr}$, the corresponding Knudsen number is $Kn\approx0.4$ for the isolated halo. Thus we expect that the ``heating" effect becomes an issue for our simulation when $Kn$ is close to $0.4$. For the SIDM subhalos, their lowest $Kn$ value ranges from $0.3\textup{--}0.6$. This may explain why their enclosed mass stalls after $t\approx4.5\textup{--}6.5~{\rm Gyr}$.

It is useful to compare to the halo presented in~\cite{Fischer:2024eaz}, where they simulated a $1.2\times10^{11}~M_\odot$ isolated halo assuming $\sigma/m=100~{\rm cm^2~g^{-1}}$. The particle mass is $3\times10^4$ and the softening length is $\epsilon=0.13~{\rm kpc}$. In that simulation, the energy starts to increase at $t\approx9.6~{\rm Gyr}$ (their Figure 1), corresponding to $Kn\approx0.1$. However, even at $Kn=0.01$ energy conservation violation is at only the $1.5\%$ level, better than our simulation. This difference is likely because~\cite{Fischer:2024eaz} used a more accurate cell-opening criterion for the gravity computations ($\texttt{ErrTolForceAcc}=5\times10^{-4}$) compared to ours ($\texttt{ErrTolForceAcc}=5\times10^{-3}$), with the former being more computationally expensive. Since the artificial heating effect leads to an underestimation of the inner density profile for a collapsed SIDM halo, our results are conservative.

\section{Convergence}
\label{sec:con}
\begin{figure}[h]
	\centering
	\includegraphics[scale=0.33]{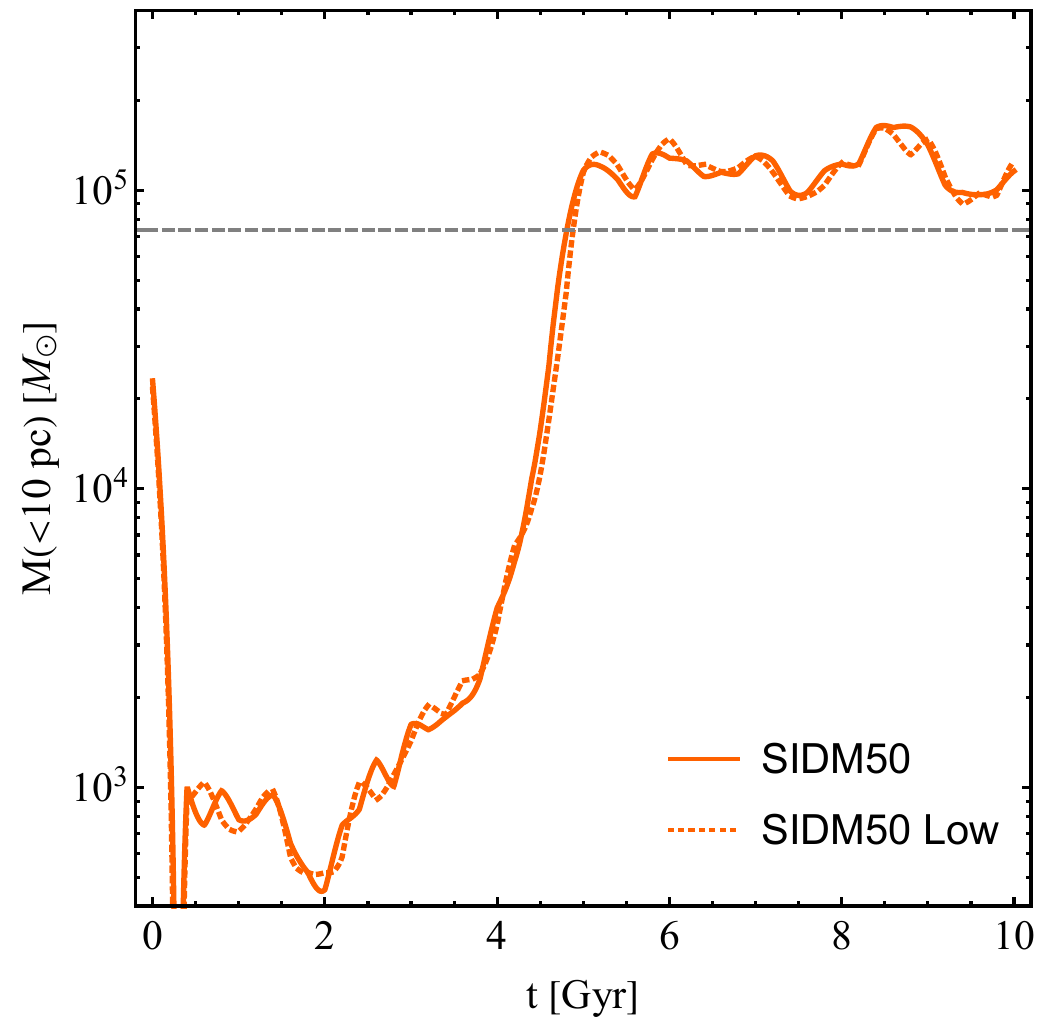}~~~
	\caption{The evolution of the enclosed mass within inner $r=10~{\rm pc}$ for the low-resolution (dotted orange) and high-resolution (solid orange) simulations.}
	\label{fig:TestMass}
\end{figure}

Lastly, to check the convergence, we have conducted an additional simulation for the SIDM50 subhalo with the total number of particles $N = 5\times10^6$, a factor of two smaller than that used to produce our main results. Figure~\ref{fig:TestMass} shows the evolution of the enclosed mass within an inner radius $r=10~{\rm pc}$ with the low-resolution (dotted orange) and high-resolution (solid orange) simulations. The results converge well. As discussed in Appendix~\ref{sec:iso}, the stalling behavior of the enclosed mass indicates that the energy conservation is violated due to the artificial heating effect. We see both high- and low-resolution simulations suffer from this issue, despite their convergence. 

\end{document}